\documentclass[journal=jacsat,manuscript=article, layout=twocolumn]{achemso}
\usepackage[version=3]{mhchem}
\usepackage{graphicx}
\usepackage{multicol}
\usepackage{xcolor}
\usepackage[normalem]{ulem}
\usepackage{xr-hyper}   % <-- use this instead of \usepackage{xr}
\usepackage{hyperref}   % must come AFTER xr-hyper
\usepackage{nameref}

\newcommand{\BaF}{BaF$_2$}
\newcommand{\PbSnSe}{Pb$_{1-x}$Sn$_x$Se}
\newcommand{\Xbar}{$\overline{\mathrm{X}}$}
\newcommand{\Gbar}{$\overline{\Gamma}$}
\newcommand{\Mbar}{$\overline{\mathrm{M}}$}
\newcommand{\Kbar}{$\overline{\mathrm{K}}$}
\newcommand{\dagga}{{\phantom{\dagger}}}
\newcommand{\g}[1]{{\bf #1}}

\author{Bartłomiej Turowski}
\affiliation{International Research Centre MagTop, Institute of Physics, Polish Academy of Sciences, Aleja Lotnikow 32/46, PL-02668 Warsaw, Poland}

\author{Wojciech Brzezicki}
\email{brzezicki@magtop.ifpan.edu.pl}
\affiliation{International Research Centre MagTop, Institute of Physics, Polish Academy of Sciences, Aleja Lotnikow 32/46, PL-02668 Warsaw, Poland}

\author{Ondřej Caha}
\affiliation{Masaryk University, Kotlářská 2, 61137 Brno, Czech Republic}

\author{Rafał Rudniewski}
\affiliation{International Research Centre MagTop, Institute of Physics, Polish Academy of Sciences, Aleja Lotnikow 32/46, PL-02668 Warsaw, Poland}

\author{Natalia Olszowska}
\affiliation{National Synchrotron Radiation Centre SOLARIS, Jagiellonian University, Czerwone Maki 98, PL-30392 Kraków, Poland}

\author{Jacek Kołodziej}
\affiliation{Faculty of Physics, Astronomy, and Applied Computer Science, Jagiellonian University, Lojasiewicza 11, PL-30348, Kraków, Poland}
\alsoaffiliation{National Synchrotron Radiation Centre SOLARIS, Jagiellonian University, Czerwone Maki 98, PL-30392 Kraków, Poland}
\affiliation{National Synchrotron Radiation Centre SOLARIS, Jagiellonian University, Czerwone Maki 98, PL-30392 Kraków, Poland}

\author{Marta Aleszkiewicz}
\affiliation{ Institute of Physics, Polish Academy of Sciences, Aleja Lotnikow 32/46, PL-02668 Warsaw, Poland}

\author{Tomasz Wojciechowski}
\affiliation{International Research Centre MagTop, Institute of Physics, Polish Academy of Sciences, Aleja Lotnikow 32/46, PL-02668 Warsaw, Poland}

\author{Tomasz Wojtowicz}
\affiliation{International Research Centre MagTop, Institute of Physics, Polish Academy of Sciences, Aleja Lotnikow 32/46, PL-02668 Warsaw, Poland}

\author{Timo Hyart}
\affiliation{International Research Centre MagTop, Institute of Physics, Polish Academy of Sciences, Aleja Lotnikow 32/46, PL-02668 Warsaw, Poland}

\author{Gunther Springholz}
\email{Gunther.Springholz@jku.at}
\affiliation{Institute of Semiconductor and Solid State Physics, Johannes Kepler University Linz, Altenbergerstr. 69, A-4040 Linz, Austria}

\author{Valentine V. Volobuev}
\email{volobuiev@magtop.ifpan.edu.pl}
\affiliation{International Research Centre MagTop, Institute of Physics, Polish Academy of Sciences, Aleja Lotnikow 32/46, PL-02668 Warsaw, Poland}
\altaffiliation{National Technical University "KhPI"{}, Kyrpychova Str. 2, 61002 Kharkiv, Ukraine}

\title[PbSnSe ARPES]
  { Tuning Dirac-Rashba and Double Dirac Cone Surface States of Topological Crystalline Insulator {\PbSnSe} by Transition Metal Adsorbate}
\begin{document}
\begin{abstract}
The electronic structure of topological insulator/magnetic metal (TI/MM) interfaces is of great importance for understanding of exotic spin-dependent phenomena and realization of advanced spin–orbitronic devices. Here, we employ a model system of submonolayer transition metal (TM) deposited on the surface of a topological crystalline insulator (TCI) of {\PbSnSe} to systematically map out the modification of the surface electronic structure by angle-resolved photoemission spectroscopy (ARPES) as a function of coverage. For the polar (111) {\PbSnSe} surface, we observe the coexistence of the Dirac topological surface states (TSS) and Rashba-split surface states (RSS) induced by the combined effects of inversion-symmetry breaking, surface band bending and orbital angular momentum effects. In particular, we demonstrate very large Rashba splittings can be obtained and the Rashba parameter ($\alpha_R$) can be tuned over a remarkably wide range from 0 to 3.5 $\text{eV}\cdot\text{\AA}$, depending on the type and coverage of the TM adatoms. Model-Hamiltonian calculations corroborate the experimental findings and reveal that this coexistence results from the filling of the TSS by the surface doping caused by the TM.  In contrast, for the nonpolar (001) surface exhibiting a double Dirac cone topological surface state, the inversion symmetry is preserved and hence no Rashba-split surface states emerge. Instead, surface charge imbalance induces dephasing of the wave functions of the double Dirac cones that diminishes the momentum-space separation between them. These findings shed light on novel phenomena occurring at the topological insulator / transition metal interface, offering a versatile platform for future spintronic and quantum devices.
\end{abstract}

\subsection{Introduction}

Topological insulators (TI) exhibit unique electronic and transport properties that are distinct from those of conventional materials. The presence of TSS offers opportunities for developing novel technologies for energy conversion and storage \cite{luo2022topological}. For example, TSS can facilitate charge separation and suppress electron–hole recombination, making TIs promising materials for photocatalytic hydrogen evolution \cite{qu2020expediting, Qu2021, xie2022progress, pan2024topological}. Moreover, owing to their helical spin texture, TIs are among the most efficient materials for spin–charge conversion \cite{Soumyanarayanan_nature_2016, Han2018_quantum_materials, wu2021magnetic, han2021topological}.

However, the influence of uncompensated magnetic moments from adjacent materials on the TSS remains unclear, and the mechanisms responsible for the large spin–charge conversion efficiency at topological insulator/magnetic metal (TI/MM) interfaces require further investigation \cite{han2021topological}. In particular, the electronic band structure at the TI/MM interface has received limited attention, despite its crucial role in determining the interconversion efficiency \cite{dey2021recent, sayed2021unified}. For instance, it has been shown that inverse Edelstein effect length, one of the main parameters characterizing interconversion, strongly depends on position of Fermi energy \cite{su2021spin}. Another example is the drastic change of Fermi contours after surface covering with rare-earth transition metals, which has been accompanied by gap opening \cite{Cano2023}. Resent experiments further demonstrated enhancement of spin–charge conversion efficiency by TI/MM interface engineering using ultrathin metallic interlayer  \cite{sun2025interface, longo2025influence}.

Monolayers of transition  metals interfaced with TIs provides possibility of time-reversal symmetry breaking and observation of quantum anomalous Hall effect (QAHE) \cite{chang2023colloquium}. This effect has great potential for quantum metrology \cite{okazaki2022quantum}, magnetic topological memristors \cite{liu2025cryogenic} and dissipationless interconnections \cite{chang2023colloquium} as well as, when combined with superconductor, may serve as a platform for fault-tolerant topological quantum computation \cite{zeng2018quantum, chang2023colloquium}. 

Among TIs, topological crystalline insulators (TCIs) are particularly attractive because their topological properties can be controlled by external parameters such as temperature, composition, crystal symmetry, thickness etc \cite{ando2015topological, liu2015crystal, Dziawa2012_PbSnSe, wagner2025probing}. TCIs based on IV-VI materials possess large spin orbit coupling (SOC) \cite{mitchell1966theoretical}, which makes them promising candidates for spin–orbitronic applications. Recently, successful spin pumping has been realised through TCI/MM interface \cite{akiyama2022direct} and a relatively large spin Hall angle of the order of 0.01 is obtained for prototype TCI SnTe at room temperature \cite{Shinobu2017spinpump}. 

Despite these advances, the TCI/MM interface remains far less explored than the TI/MM interface.  It is worth noting that the band structure evolution at the TCI/MM interface has not been given significant consideration despite great opportunity to control of TSS and correspondingly magnetic properties of adjacent MM by external means. Moreover, it has been recently shown that in addition to TSS, giant Rashba-split states \cite{volobuev2017giant, rechcinski2021rashba} can arise in this material. The presence of both TSS and Rashba split states may result in significant enhancement of spin-charge conversion parameters \cite{sun2019large, tong2020enhanced, Shi2018efficient} and enable their control via bias current and gate voltage \cite{hoque2024room} as shown in recent experimental works. Magnetic dopants on the TCI surface are also of fundamental interest, as they can induce two-dimensional surface magnetism through RKKY interactions \cite{reja2017surface, FERTIG2019113623, RejaSpinStiffness, klier2019tuning, pankratov2019understanding, hoi2024switching}.

Here, we present a combined experimental and theoretical investigation of TCI {\PbSnSe} epilayers with submonolayer coverages of Fe and Mn transition-metal (TM) adsorbates. The high-quality epilayers were grown using molecular beam epitaxy (MBE) in both (111) and (001) orientations and subsequently transported to the synchrotron under ultra-high vacuum (UHV) conditions or covered with protective layer for in-situ metal deposition and ARPES measurements.

For the conduction band of (111)-oriented surfaces, alongside the TSS we detect spin-split electronic bands, indicating a strong Rashba-type interaction whose strength varies with the transition-metal coverage. In contrast, on the (001) surfaces, the deposition does not give rise to Rashba splitting but instead alters the overlap of the double Dirac cones, leading to a noticeable contraction of their momentum-space separation. These results demonstrate that even ultrathin magnetic overlayers can profoundly influence the surface band structure of TCIs, providing a model platform for investigating interfacial spin–orbit coupling and magnetic proximity effects.
%{\color{red} In the case of (111)-oriented samples, the emergence of Rashba split surface states (RSS) in the conduction band was observed. The estimated Rashba parameter ($\alpha_R$) exhibited remarkable tunability, ranging from 0 to 3.5 eV·Å, depending on the type of deposited TM.For the (001)-oriented films, our observations revealed a reduction in the separation of the Dirac points within the double Dirac cone in k-space (Fig. 1b). The modifications in the band structure resulting from the TM deposition will be thoroughly discussed, and potential reasons for these alterations will be addressed. }

\begin{figure*}[h]
    \includegraphics[width=\columnwidth*2]{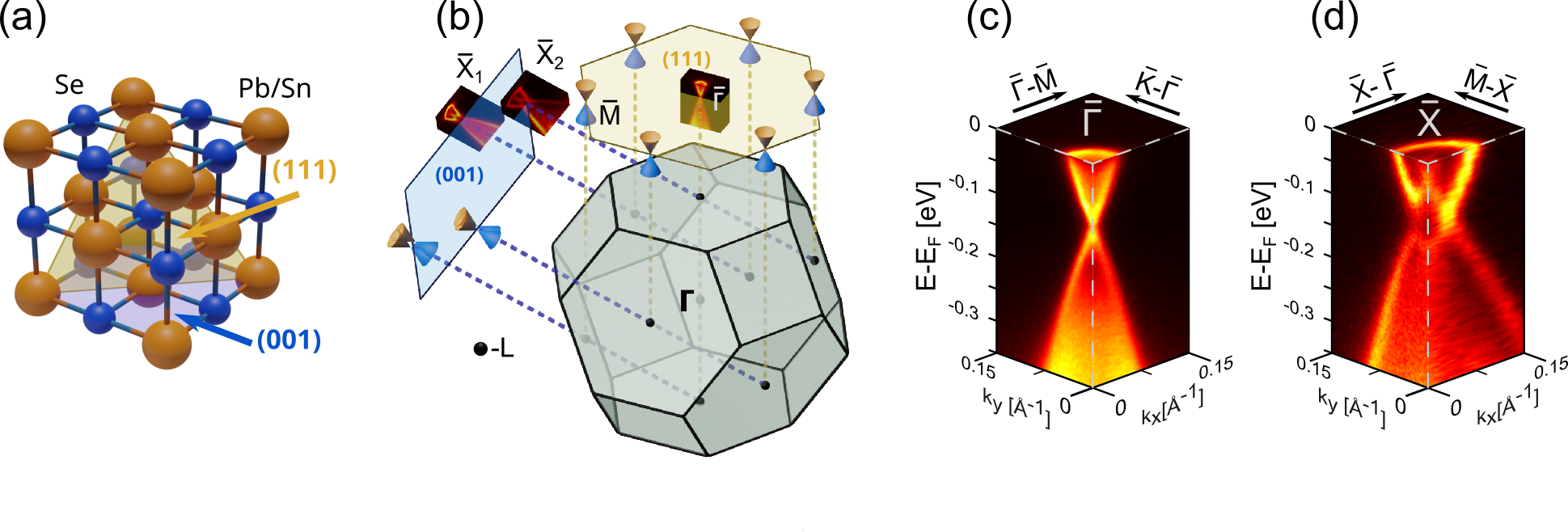}
    \caption{ Crystal and electronic structure of {\PbSnSe} TCI. (a) Unit cell of {\PbSnSe} with (111) and (001) planes highlighted. (b) Schematic bulk and surface Brillouin zones (BZs) corresponding to the (111) and (001) orientations, with high-symmetry points and projected locations of the Dirac cones indicated. (c,d) Three-dimensional ARPES intensity maps measured at 80~K in vicinity of {\Gbar} and {\Xbar} points for Pb$_{0.75}$Sn$_{0.25}$Se epilayers grown on BaF$_2$ (111) and KCl (001) respectively. The corresponding places at surface BZs are indicated in panel (b).}
    \label{fig:figure1}
\end{figure*}

%In TCI phase lead tin selenide (PbSnSe) has rock salt structure (Figure \ref{fig:figure1} (a)) and  mirror symmetry of (110) plane protects the TSS present on (111) and (001) planes. L points of bulk Brillouin zone project on (001) plane as double Dirac cones (DC) in {\Xbar} points and on (111) plane in {\Gbar} and {\Mbar} points of surface Brillouin zones (Figure \ref{fig:figure1} (b)). Submonolayer deposition of transition metals, schematically shown in the Figure \ref{fig:figure1} (c), causes evolution of band structure, which may be probed by angle-resolved photoemission spectroscopy (ARPES), as presented in Figure \ref{fig:figure1} (d) and (e).

\section{Methods}
\label{aaa}

\subsection{ Experimental}
It is well-established that {\PbSnSe}, undergoes a topological phase transition from a normal insulator to a topological crystalline insulator (TCI) at specific Sn concentrations and temperatures \cite{Nimtz_kp, Dziawa2012_PbSnSe, krizman2018topphasediagr}. In this work, we focus on the Sn content and temperature ranges corresponding to the TCI phase only. In the TCI phase, {\PbSnSe} crystallizes in a rock salt structure (Figure \ref{fig:figure1} (a,b)), where the mirror symmetry of (110) planes protects the TSS present on the (111) and (001) surfaces. The L points of bulk Brillouin zone project on (001) surface as double Dirac cones (DC) in {\Xbar} points and on (111) plane in {\Gbar} and {\Mbar} points of surface Brillouin zones (Figure \ref{fig:figure1} (b)). Submonolayer deposition of transition metals causes evolution of band structure, which may be probed by ARPES, yielding spectra similar to those shown in Figure \ref{fig:figure1} (c) and (d).

Two sets of 1-$\mu$m thick {\PbSnSe} epilayers with $x_{Sn}$= 0.25 and 0.3 were grown on (111) {\BaF} and (001) KCl substrates (Figure \ref{fig:figure1}~(c,d))  in Riber 1000 and Veeco GENxplor molecular beam epitaxy (MBE) systems, under growth condition similar to those reported previously  \cite{assaf2017magnetooptical,mandal2017topological, bermejo2023observation}. For these Sn concentrations, the band gap is closed by topological surface states (see Figure \ref{fig:figure1} (c,d) and Figure S5), as observed at 80~K in agreement with ref.\cite{krizman2018topphasediagr,Nimtz_kp}. Moderate Bi doping (typically below 0.1 at. \%) was used to tune Fermi level to desired position \cite{mandal2017topological,volobuev2017giant}. More details on samples growth and characterization can be found in Section S1, Section S2, Section S3, Section S4.

The first set of epilayers was covered \textit{in-situ} in the MBE chamber with an amorphous Se layer and later decapped in ARPES preparation chamber by Se reevaporation. The second set was transported to ARPES system in Ferrovac VSN40S ultra-high vacuum (UHV) suitcase\cite{volobuev2017giant}, without breaking vacuum. Both sample sets (Se-capped and UHV-transferred) exhibited similar ARPES spectra making obtained results self-consistent and reproducible.   

Transition metals, specifically Mn and Fe, were deposited on a clean {\PbSnSe} surface at room temperature. The deposition was performed in preparation chamber of ARPES endstation at URANOS (former UARPES) beamline of SOLARIS synchrotron (Krakow)\cite{olszowska2026uranos} with incremental steps of 0.05-0.2 monolayer (ML). After each deposition, the samples were transferred to analytical chamber, where E(k) ARPES spectra were recorded.

The ARPES investigations were performed using horizontally polarized light with photon energies in the range of 15-90 eV, near {\Gbar} point and {\Xbar} points of surface Brillouin zone for (111)- and (001)-oriented films respectively. Results obtained near the {\Mbar} point (see Figure S6) were not discussed here due to relatively weak spectral intensity at this point. High-quality spectra were acquired using a Scienta Omicron DA30L photoelectron spectrometer with minimum energy resolution of 1.8 meV and angular resolution of 0.1$^{\circ}$.  The best resolution and optimal visibility of the spectra were achieved at photon energies of $17-19$ eV and temperature 80K as previously reported\cite{Dziawa2012_PbSnSe,mandal2017topological,rechcinski2021rashba}. Core level (CL) spectra (Figure S10) were recorded at 90eV to determine surface elemental composition of the epilayers. 

\subsection{Theoretical calculations}

%Our starting point is the tight-binding (TB) Hamiltonian for SnTe-material class \cite{Hsieh2012}, 
Our theoretical description is based on the tight-binding (TB) model developed for the SnTe family of topological crystalline insulators (TCIs) by Hsieh et al. \cite{Hsieh2012}, to which \PbSnSe~belongs,
\begin{eqnarray}
H&=&m\sum_j (-1)^j\sum_{\g r, \alpha} \hat c^\dagger_{j\alpha}(\g r)\cdot \hat c^\dagga_{j\alpha}(\g r)\nonumber\\
&+& \sum_{j,j'}t_{jj'}\sum_{\langle\g r,\g{r'}\rangle,\alpha} \hat c^\dagger_{j\alpha}(\g r)\cdot \hat d_{\g r\g{r'}}\ \hat d_{\g r\g{r'}}\cdot \hat c^\dagga_{j'\alpha}(\g{r'})\nonumber\\
&-&\sum_j i\lambda \sum_{\g r,\alpha,\beta}\hat c^\dagger_{j\alpha}(\g r)\times \hat c^\dagga_{j\beta}(\g r)\cdot \hat \sigma_{\alpha,\beta},
\label{eq:tbmodel}
\end{eqnarray}
where $\hat c_{j \alpha}(\mathbf{r})$ are vectors of fermionic operators corresponding to $p_x$-, $p_y$- and $p_z$-orbitals and the indices denote the sublattice $j\in\{1,2\}$ [(Sn,Pb)/Se atoms], spin $\alpha$ and lattice site $\mathbf{r}$. Here $\hat \sigma_{\alpha, \beta}$ is a vector of Pauli matrices, $\hat d_{\g r\g{r'}}$ are unit vectors pointing from $\g r$ to $\g{r'}$ and the next-nearest-neighbour hoppings satisfy $t_{11}=-t_{22}$.
Defining the unit cell as two atoms at positions $(0,0,0)$ and $(0,0,1)$ (taking one interatomic distance as a length unit) and the lattice translation vectors as $\vec{a}_1=(1,0,1)$, $\vec{a}_2=(0,1,1)$ and $\vec{a}_3=(0,0,2)$ we find that the three-dimensional bulk Hamiltonian can be represented in momentum space, \cite{Brzezicki2019} as presented in Eq. (Seq2).

The presence of the metallic layers at the $(111)$ surface is modeled in the following way: we assume that the  effect of the deposition is to fill in the surface states with charge. We assume that they are filled up to momenta $k=\pm1/\xi$ and having their spatial charge density from the diagonalization of the TB Hamiltonian we solve the Poisson equation to obtain the shape of the induced surface potential $V(r_{\perp})$ as function of the distance from the surface $r_{\perp}$. This electrostatic potential is further modified by the Thomas-Fermi screening, so the final formula takes the form of $V_{\rm sc}(r_{\perp})=\eta V(r_{\perp})e^{-r_{\perp}/\lambda_{\rm TF}}$, where $\lambda_{\rm TF}$ is the Thomas-Fermi screening length and $\eta$ is the free parameter related to the total amount of charge that flows into the surface states.
 
\section{Results}

\subsection{Fe deposition}

\begin{figure*}[!ht]
    \includegraphics[width=\columnwidth*2]{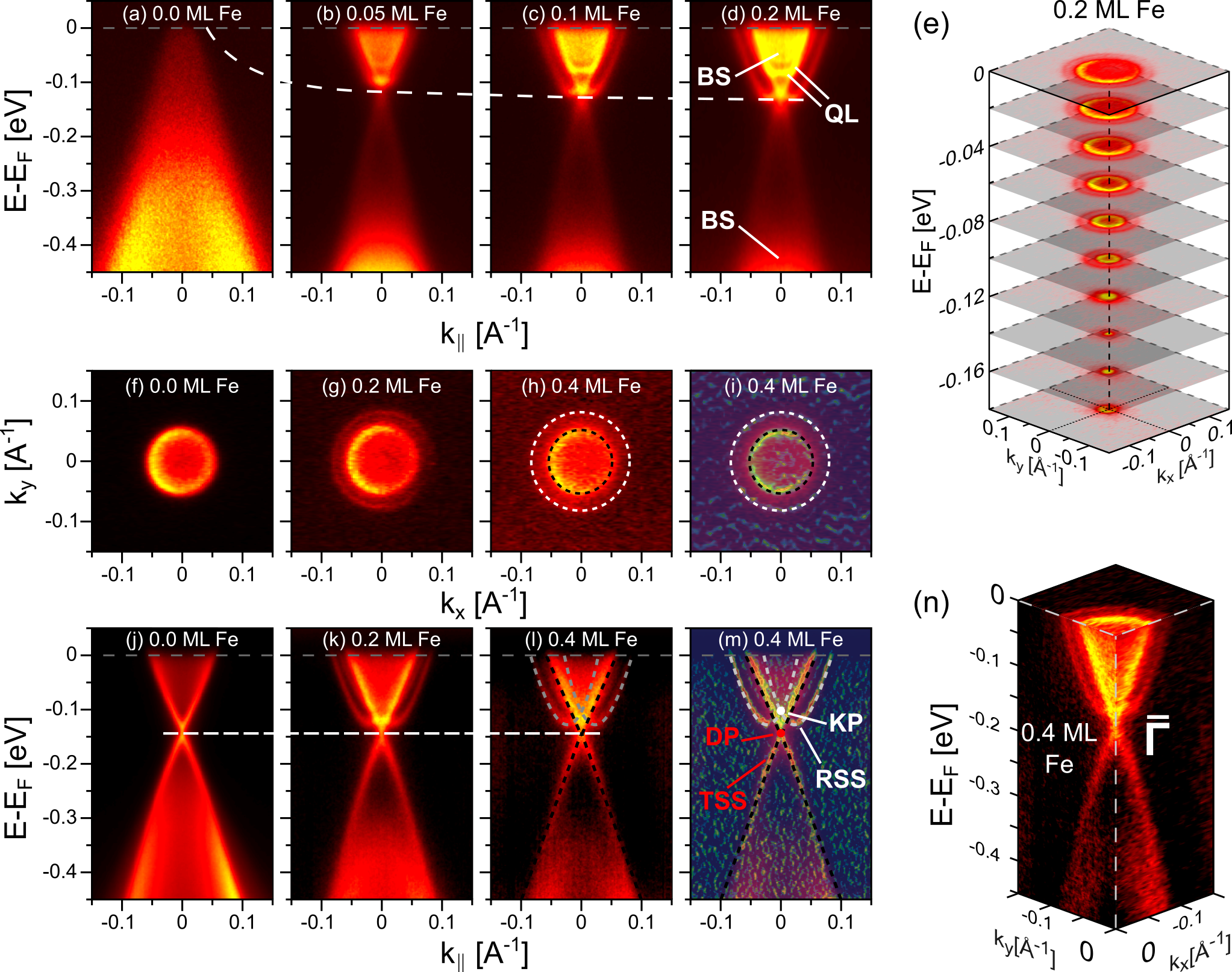}
    \caption{ Evolution of the electronic band structure of (111)-oriented {\PbSnSe} upon Fe deposition. (a-e) ARPES spectra of p-type Pb$_{0.75}$Sn$_{0.25}$Se recorded near the  {\Gbar}  point at 80 K with photon energy of 18 eV, showing progressive changes in the surface band structure with increasing Fe coverage. The white dashed lines indicate the position of the Dirac point (DP). Bulk states (BS) and quantized quantum well levels (QL) originating from band bending are marked in panel (d). (f–n) Corresponding ARPES spectra of n-type Pb$_{0.7}$Sn$_{0.3}$Se under the same conditions. Fermi surface maps (f–h) and constant-energy maps (e) reveal the emergence of Rashba-split surface states (RSS), which overlap with the topological surface states (TSS) in the conduction band. (i,m) 2D curvature plots highlighting the Rashba-type dispersion of the surface states. The Kramers point (KP) and Dirac point (DP) denote the crossing positions of the RSS and TSS, respectively. (n)~Three-dimensional ARPES intensity map illustrating the full Rashba-split band structure.}
    \label{fig:figure2}
\end{figure*}

Figure \ref{fig:figure2} shows the evolution of the electronic spectra upon Fe deposition on the (111) surface of {\PbSnSe} and also compares two samples: a p-type film (Figure \ref{fig:figure2} (a-e)) and an n-type film with stronger Bi doping (Figure \ref{fig:figure2} (f-n)). Although the samples have slightly different composition ($x_{Sn}$=0.25 and 0.3 respectively), they demonstrate similar Dirac-like energy dispersion with crossed TSS in pristine state (Figure \ref{fig:figure2} (a,j)). The Fermi energy ($E_F$) lies in valence (VB) and conduction band (CB) for p- and n-type samples respectively. Broad bands from bulk states are also visible inside TSS cone and marked in the Figure \ref{fig:figure2} (d) as BS. The TSS exhibit a circular Fermi surface (Figure \ref{fig:figure2} (e-i)), making the spectrum in {\Gbar}-{\Mbar} and {\Gbar}-{\Kbar} directions equivalent in vicinity of {\Gbar} point. Upon progressive submonolayer Fe deposition, new states separated in momentum space emerge in CB. The states arrange in concentric circles in Rashba like manner, as it is evident from the Fermi surface plots (Figure \ref{fig:figure2} (g-i)) and constant energy surface maps (Figure \ref{fig:figure2} (e)). The Rashba-split surface states (RSS), which overlap with TSS in the CB, are most clearly visible in 3D representation of electronic structure (Figure \ref{fig:figure2} (n)). 

The intersection of the RSS branches appears at Kramers point (KP), located at or slightly above the Dirac point (DP), where the TSS cross Figure \ref{fig:figure2} (m). The inner RSS branches overlap with TSS in the CB. As seen in Figure \ref{fig:figure2}, the Rashba splitting increases with Fe coverage, whereas the behavior of $E_F$ differs for the n-doped an p-doped samples. In the case of p-type PbSnSe, a pronounced upward shift of $E_F$ occurs after deposition as little as 0.05 ML Fe (Figure \ref{fig:figure2} (a,b)). Further Fe deposition has a weaker effect on $E_F$, as illustrated by the white dashed line marking the shift in the position of the Dirac points. In contrast, $E_F$ remains essentially unchanged for the n-doped sample (see Figure \ref{fig:figure2} (j-l)). 

In addition to the Rashba-split states, well-resolved quantum levels (QLs) arising from a two-dimensional electron gas (2DEG) are observed (Figure \ref{fig:figure2} (d)). Surface Fe deposition induces downward band bending, leading to the formation of an asymmetric quantum well \cite{chen2012robustness,bianchi2012robust,bianchi2011simultaneous,bahramy2012emergent} (schematically shown in Figure \ref{fig:figure3} (f)).  

To visualize the surface state dispersion more precisely, the 2D curvature method\cite{Pan_curvature2D} was applied (Figure \ref{fig:figure2} (i,m)). The Rashba spectrum is fitted using following dispersion relation taking into account second order correction \cite{soumyanarayanan2015momentum}: 
\begin{equation}
E_\pm = \varepsilon_0+\frac{\hbar^2k^2}{2m} \pm \alpha_R (1+\nu k^2)k
\label{eqn:dispersionRelation}
\end{equation}
where $\varepsilon_0$ - the band-offset energy, $m$ - the effective mass of the electrons, $k$ - in-plane electron momentum, $\nu$ - coefficient describing the second-order correction and $\alpha_R$ is the Rashba parameter extracted from the fitting procedure. The results of the fitting is presented in Figure \ref{fig:figure2} (h,i,l,m).
   
\subsection{Mn deposition}

\begin{figure*}[!h]
    \includegraphics[width=\columnwidth*2]{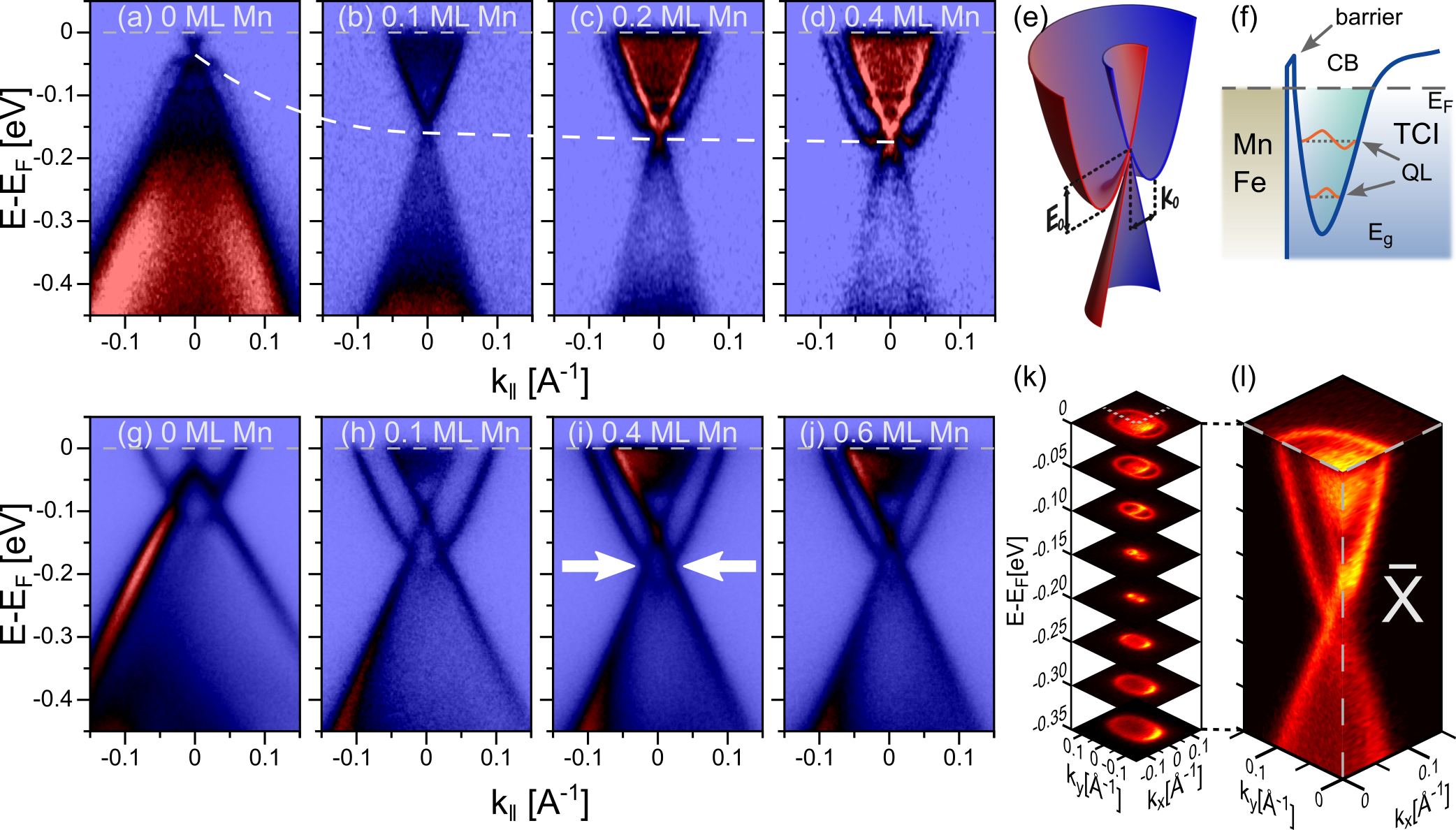}
    \caption{Evolution of the electronic band structure of \PbSnSe\ upon Mn deposition. (a–d) ARPES spectra of (111)-oriented Pb$_{0.7}$Sn$_{0.3}$Se/BaF$_2$ recorded near the {\Gbar} point at 80 K with photon energy of 18 eV, showing systematic modifications of the surface band structure with increasing Mn coverage. The dashed lines trace the shift of the chemical potential, defined by the position of the DP. (e) Schematic representation of the observed spectra, illustrating the formation of Dirac–Rashba SS. (f) Schematic diagram showing the development of quantized quantum-well (QL) states near the surface of \PbSnSe\ induced by Mn or Fe deposition.  (g–j) ARPES spectra of (001)-oriented Pb$_{0.75}$Sn$_{0.25}$Se/KCl recorded near the {\Xbar} point at 8 K with photon energy of 18 eV, revealing a progressive reduction of the in-plane momentum separation between the double DCs with increasing Mn coverage. (k) Constant-energy surface maps and (l) three-dimensional ARPES intensity map for 0.2 ML Mn deposited on the (001) surface, highlighting the double-Dirac-cone features.}
    \label{fig:figure3}
\end{figure*}

To examine whether other transition metals produce similar effects on the band structure of \PbSnSe, Mn was deposited on both the (111) and (001) surfaces. For submonolayer Mn coverage, a comparable evolution of band bending and the emergence of RSS were observed on the (111) surface (Figure \ref{fig:figure3} (a-d)). In pristine state (Figure \ref{fig:figure3} (a)), moderate Bi doping was employed to position the Fermi level slightly above the Dirac point. After deposition of 0.1 ML of Mn (Figure \ref{fig:figure3} (b)), a pronounced upward shift of the Fermi energy by approximately 100 meV was detected, and well-defined TSS of the conduction band became visible. Further deposition of 0.2 ML of Mn (Figure \ref{fig:figure3} (c)) revealed a clear Rashba splitting, while increasing the coverage to 0.4 ML (Figure \ref{fig:figure3}~(d)) resulted not only in pronounced Rashsba SS, but also in well-resolved  QW levels (schematically illustrated in Figure 3 (f)). The crossing of the RSS and TSS occurs nearly at the same energy and momentum, forming a characteristic Dirac–Rashba spectrum (Figure \ref{fig:figure3} (e)), similar to that observed for Fe-deposited surfaces (Figure \ref{fig:figure2}). 

It should be noted that even in the pristine state, weak RSS features are visible in the valence band, which may result from residual surface adsorbates \cite{king2011large, benia2011reactive, bianchi2011simultaneous} or from excess Se on the surface. \cite{rechcinski2021rashba}. The latter explanation is more likely, given that the pristine surfaces in this study were prepared by Se re-evaporation in the Solaris synchrotron preparation chamber. 

Although the (001) surface exhibits a similar upward shift in the Fermi level upon Mn deposition, Rashba splitting is absent. Instead, a reduction in the momentum-space separation between the Dirac points (DPs) of the double Dirac cones (DCs) is observed (Figure \ref{fig:figure3} (g-j)). The most significant upward shift in E$_F$, accompanied by the strongest reduction in DP separation, occurs after the deposition of 0.1 ML of Mn (Figure \ref{fig:figure3} (h)). Increasing the coverage to 0.4 ML and 0.6 ML leads to further slight E$_F$ shifts and a continued decrease in the double-DC separation (Figure \ref{fig:figure3} (i-j)). Despite these quantitative changes, the overall band structure remains qualitatively the same, as shown in (Figure \ref{fig:figure3} (k-l)). The two DPs of the double DCs remain clearly resolved in both constant-energy surface maps and 3D ARPES spectra, and no Lifshitz transition \cite{pletikosic2014inducing, neupane2015topological} is detected.    

\subsection{Theory}

\begin{figure}[h]
    \includegraphics[width=\columnwidth]{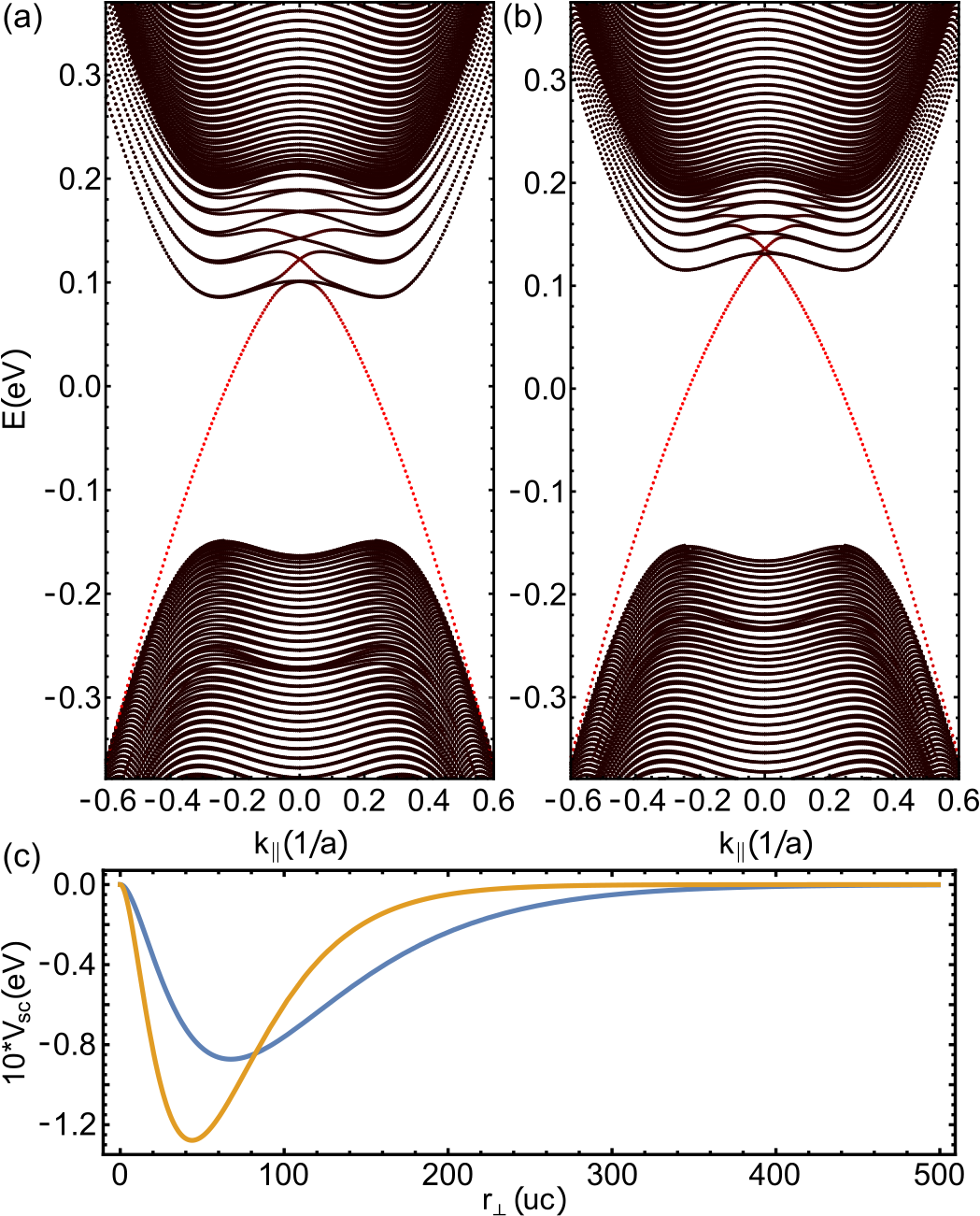}
    \caption{Energy bands of the TB model for slab with $(111)$ surfaces in presence of the surface potential  with $\xi = 10$ and: (a) $\eta=0.24$, $\lambda_{\rm TF}=30$, (b) $\eta=0.08$, $\lambda_{\rm TF}=50$ (all lengths in unit cells). Red color lines indicate TSS on a top surface of the slab, states localized at bottom surface were removed. (c) Shape of the potential as a function of distance $r_{\perp}$ from the surface. The yellow and blue curves correspond to the parameters used in (a) and (b), respectively.}
    \label{fig:figure_th1}
\end{figure}

In Fig. \ref{fig:figure_th1} we show the modeled band structure of the slab with $(111)$ surfaces in the presence of a surface potential generated by
charge filling up the TSS. The spectra are obtained for $N=500$ layers of primitive unit cells using
the tight-binding model defined by Eq. (\ref{eq:tbmodel}) with microscopic parameters set as: $m=2.2$, $\lambda =0.5$,
$t_{12}=2$, $t_{11}=0.56$ (all in eV). Such choice of parameters assures that: ($i$) the model is in the TCI phase, ($ii$) 
bulk gap and surface Dirac cones resemble ones observed experimentally. Our goal here is not to reproduce all the
details of the band structure but rather to provide a qualitative explanation of the effect. The spectra calculated
for two representative choices of the surface potential parameters $\eta$ and $\lambda_{\rm TF}$ indeed prove that 
TSS can coexist with RSS, provided that the minimum of the surface potential is located at a sufficient distance from the surface itself. See the Figure S11 for the evolution of the spectra
with increasing magnitude $\eta$ of the potential.

\section{Discussion}

The effect of surface doping of topological materials with TMs carrying uncompensated magnetic moments has been studied experimentally in several works \cite{valla2012photoemission, scholz2012tolerance, wang2015robust,TUROWSKI2023}. However, the TSS remain robust and gapless, rather than developing the expected magnetic gap\cite{Marom2020SnSeEuSe, wangassaf2023energy, majewski2026many}. A similar behavior is observed in our TCI samples. Despite recent theoretical reports predicting ferromagnetic (FM) ordering on the surface of TCIs \cite{reja2017surface, RejaSpinStiffness,FERTIG2019113623}, and consequently a gap opening at the Dirac point, our measurements in the 8–80~K range (well below the topological phase transition temperature) reveal undisturbed, gapless TSS within experimental uncertainty of $<$ 10~meV. Apparently, the observation of 2D FM states in TCI requires precise tuning of $E_F$ to DP, which is experimentaly challenging. Another limiting factor could be the extremely low Curie temperature of such surface-localized FM states, lying below the temperature range accessible in our current experiments. Thus, the realization of a ferromagnetism-induced gap in TCI surface states remains experimentally elusive.

One of the most important findings of this work is the coexistence of Dirac and Rashba surface states, which we call Dirac-Rashba (Figure \ref{fig:figure3} (e)). Although both TSS and RSS have been reported previously in various topological materials, they typically appear as separated features both in energy and momentum\cite{shoman2015topological}.

The first theoretical prediction of Dirac-Rashba hybrid states was made for BiTeI trilayer on top of the TI PbSb$_2$Te$_4$ \cite{eremeev2015new}.  Later, the concept of Dirac-Rashba states was extended to graphene-based systems with strong spin-orbit coupling (SOC) \cite{milletari2017covariant, khokhriakov2020gate}, and more recently to topological superlattices consisting of alternating quintuple/septuple layers \cite{shvets2024interplay}. The latest theoretical work by Cuono et al. \cite{Cuono25} addressed the coexistence of Rashba and Dirac surface states in centrosymmetric TCIs decorated with metalic overlayers. Their calculations predict that the Rashba-split states originate primarily from subsurface layers, whereas the Dirac surface states remain localized in the outermost atomic layers. This spatial separation minimizes the overlap of their wavefunctions, leading to coexistence without hybridization. Our present theory agrees with this view and provides more intuitive picture base of a model-Hamiltonian approach. This is also more flexible as it allows us to tune the parameters of the surface potential, include Thomas-Fermi screening and relate its origin to a concrete physical mechanism, namely charge from TM adsorbates filling in the TSS.
 
 %Our ARPES data on Pb$_{1-x}$Sn$_x$Se support this scenario: the RSS and TSS appear within the same energy–momentum range and remain largely decoupled, exhibiting only subtle spectral-weight redistribution consistent with weak coupling. 
 Experimentally, Dirac-Rashba states have been observed in other TCI systems such as (111) Pb$_{1-x}$Sn$_x$Te heavily doped with Bi \cite{volobuev2017giant}, and in asymmetric quantum wells of {\PbSnSe} grown on (111) Pb$_{1-x}$Eu$_x$Se \cite{rechcinski2021rashba}. More recently, similar Dirac-Rashba-like coupled spectra have been detected for alkali-metal deposited on topological Dirac semimetal (001) KZnBi \cite{lee2024progressive}. In all of these systems, the observed split states were explained within the Bychkov–Rashba model \cite{bychkov1984properties}, in which the spin splitting originates from inversion-symmetry breaking and the associated surface electrostatic potential. 
Our results are consistent with this picture. However, we find that the inner branches of the Rashba-split surface states remain pinned to the topological surface states over a broad range of transition-metal coverages, with their crossing position remaining essentially unchanged despite the increasing Rashba splitting. Such a robust Dirac–Rashba spectrum is particularly attractive for spintronic applications, as it provides a versatile platform for tuning and enhancing spin–charge interconversion efficiency. \cite{sun2019large, Wang_bilinear}. 

\begin{figure}[!h]
    \includegraphics[width=\columnwidth]{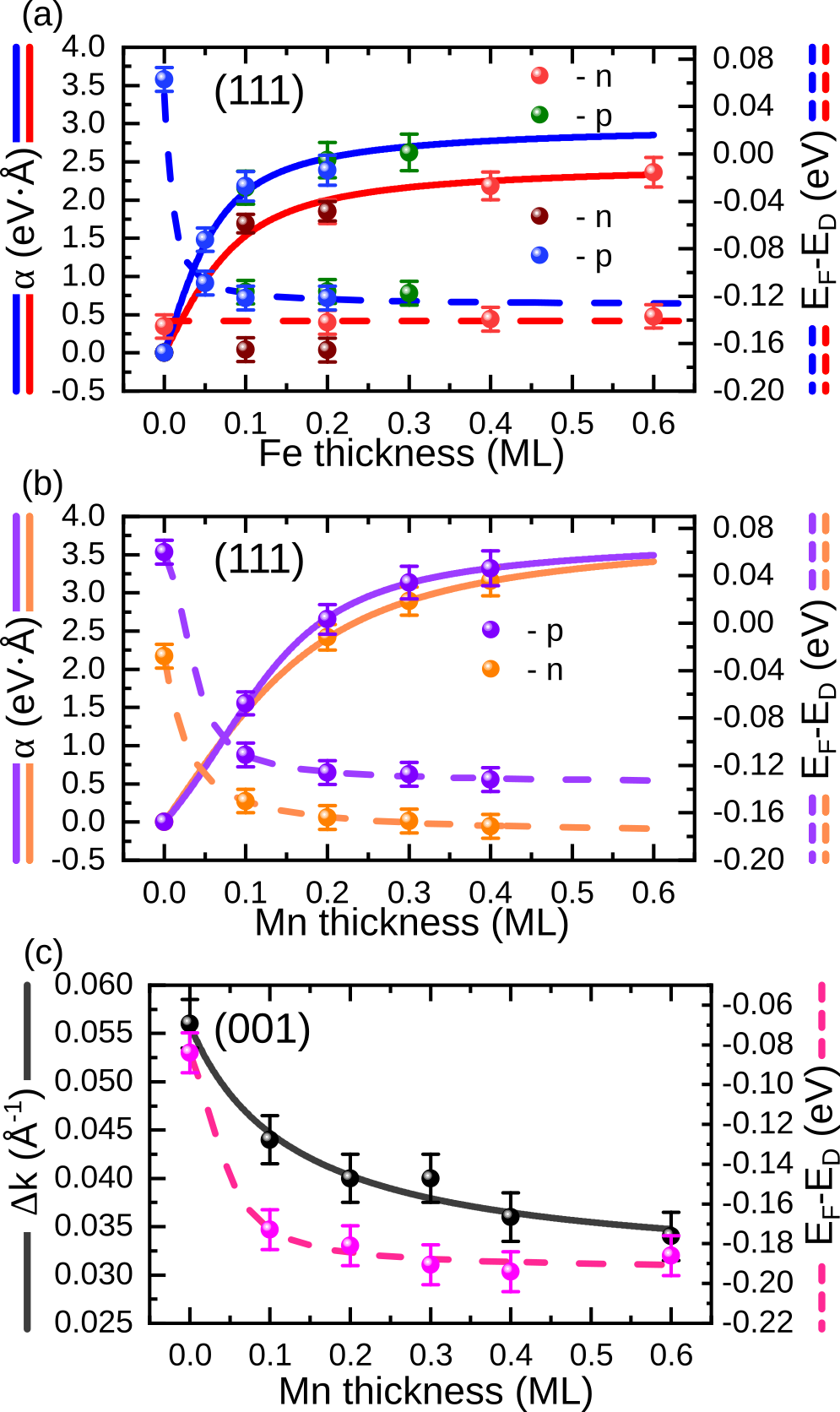}
    \caption{Influence of submonolayer TM deposition on (111) and (001) surfaces of TCI {\PbSnSe}. (a,b) Variation of the Rashba parameter $\alpha_R$ (solid lines) and the Fermi level shift ($E_F-E_D$) (dashed lines) as functions of Fe (a) and Mn (b) submonolayer thickness for the (111)-oriented surface. Deposition of both Fe and Mn leads to an initial increase, followed by saturation of  $\alpha_R$  for a coating thickness of 0.2–0.4 ML. (c) Dependence of the momentum separation between the double Dirac cones (solid line) and the corresponding ($E_F-E_D$) (dashed lines) on Mn coverage for n- and p-type surfaces of the (001) films. The lines are guides for the eye.}
    \label{fig:figure4}
\end{figure}

The main experimental results for (111)-oriented epilayers are summarized in Figure \ref{fig:figure4} (a,b) which show the fitted Rashba parameter ($\alpha_R$) and determined Fermi level relative to the Dirac point ($E_F-E_D$) as a function of TM coverage. For both Mn and Fe, $\alpha_R$ increases with coverage and saturates for  0.2-0.4 ML, reaching values of 2.5 - 3.5~eV·\AA, which are typical for other giant Rashba-split systems \cite{AstGiantSpinSpliting, ishizaka2011giant}. One may notice that the position of $E_F$ follows a trend correlated with $\alpha_R$, similar to previous reports \cite{volobuev2017giant, rechcinski2021rashba}, although $E_F$ reaches the saturated value more rapidly than $\alpha_R$. This suggests that factors beyond electrostatic band bending contribute to the large Rashba splitting. Indeed, in the heavily n-doped sample (see Figure \ref{fig:figure4} (a)), where saturation of $E_F$ is already achieved, $\alpha_R$ continues to increase with further TM deposition, supporting the notion that the giant Rashba effect cannot be explained solely by surface potential gradients. 

Recent works have shown that the interface electric field alone is insufficient to produce such large Rashba splitting. Instead, the combined effects of atomic spin–orbit coupling (SOC) and the interplay of orbit angular momentum (OAM) \cite{park2011orbital} with surface electrostatic potential result in such a giant effect \cite{unzelmann2020orbital, bihlmayer2022rashba}. The presence of OAM in our system is consistent with the observed strong circular dichroism in the ARPES spectra (Figure S9) that are highly sensitive to the orbital texture \cite{boban2025scattering}. Additionally, the presence of strong SOC is confirmed by the pronounced weak antilocalization effect observed in transport measurements. \cite{kazakov2021signatures, kazakov2025topological}. Therefore, we suggest that the observed giant Rashba splitting in {\PbSnSe} is not only governed by the surface potential and a strong SOC, but also receives enhancement from the interplay with OAM. Specifically, adsorption of transition metals can locally modify the orbital character (e.g. increase mixing of p and d orbitals) at the surface or near-surface layers, thereby boosting OAM contributions to the Rashba effect and amplifying SOC-derived spin splitting. Such a mechanism can lead to an enhanced Rashba coefficient even when the Fermi level reaches saturation, and helps to explain why $\alpha_R$ continues to grow in some of our samples despite limited further band bending.  

For the (001)-oriented epilayers, Figure \ref{fig:figure4} (c) shows the dependence of the double-Dirac-cone separation and $E_F-E_D$ on Mn coverege. Similar to the (111)-oriented films, we observe a reduction of the momentum separation between the Dirac cones accompanied by a shift  of $E_F$. However, no Rashba splitting is detected for this orientation, as the (001) surface preserves most of the bulk crystal symmetries. In this case, alternating cations and anions at the surface balance the electrostatic potential, maintaining inversion-like symmetry even in the presence of TM-induced fields, which suppresses the Rashba effect. Nevertheless, surface perturbations introduced by TM adsorption lead to dephasing of the wavefunctions associated with the double Dirac cones, resulting in the observed reduction of their k-space separation.{\cite{polley2018fragility}. 

Interestingly, (001)-oriented films exhibit a helical spin texture {\cite{wojek2013spin} similar to that of the Dirac–Rashba states on the (111) surface, however in this case both inner and outer surface states Dirac cones should be topologically protected. Such a configuration should also support tunable spin–charge interconversion, although the overall splitting is expected to be suppressed by transition-metal deposition.

\section{Conclusion}

 Submonolayer deposition of transition metals on {\PbSnSe} topological crystalline insulators produces orientation-dependent modifications of the surface electronic structure. On the polar (111) surface, Fe and Mn induce Rashba-split surface states that weakly couple with the topological surface states, forming Dirac–Rashba spectra with tunable Rashba parameter reaching a value of up to 3.5 eV·Å. This giant splitting arises not only from surface potential gradients but also from the interplay of atomic spin–orbit coupling and orbital angular momentum. In contrast, the non-polar (001) surface preserves inversion symmetry and shows no Rashba splitting, but instead exhibits a reduction of double Dirac cone separation due to charge imbalance and wavefunction dephasing. For both orientations, the topological surface states remain gapless after the deposition of magnetic atoms, which highlights the robustness of TCI surface states and the experimental challenges inducing magnetic gaps. These findings establish the transition metal/TCI interface as an effective tool for engineering Rashba coupling and Dirac cone dynamics in TCIs, with direct implications for spin–orbitronic applications. %Based on our recent measurements we also predict that similar effects of band structrure change can be induced by other metals including superconducting Pb and Sn splitting will be observed in other (111) IV-VI semiconductors and TCI including PbTe. This is of particular interest for current development superconductor/IV-VI semiconductor qubits.} 

\begin{acknowledgement}

The authors thank G. Bauer, T. Dietl, T. Story and R. Buczko for valuable discussions. The authors acknowledge the use of ARPES 3D-map visualization scripts developed by M. Rosmus. The
PbSe and SnSe binary compounds used as source materials for the effusion cells were prepared in the Institute of Physics Polish Academy of Sciences by J. Korczak. This research was partially supported by the Foundation for Polish Science project "MagTop" no. FENG.02.01-IP.05-0028/23, co-financed by the European Union from the funds of Priority 2 of the European Funds for a Smart Economy Program 2021-2027 (FENG).  V.V.V. acknowledges long-term program of support of the Ukrainian research teams at the Polish Academy of Sciences carried out in collaboration with the U.S. National Academy of Sciences with the financial support of external partners and Narodowe Centrum Nauki (NCN, National Science Centre, Poland) IMPRESS-U Project No. 2023/05/Y/ST3/00191. W.B. acknowledges support from Narodowe Centrum Nauki (NCN, National Science Centre, Poland) Project No. 2019/34/E/ST3/00404. OC was supported by QM4ST project financed by the Ministry of Education of Czech Republic, grant no. CZ.02.01.01/00/22\_008/0004572. GS thanks for the support Austrian Science Fund (Grant No. AI0656811/21) and the LIT Grant No. LIT-2022-11-SEE-131 of the University of Linz.
Publication subsidized  under the provision of the Polish Ministry of Science and Higher Education project "Support for research and development with the use of research infrastructure of the National Synchrotron Radiation Centre SOLARIS” under contract nr 1/SOL/2021/2. We acknowledge SOLARIS Centre for the access to the Beamline URANOS (former UARPES) where the measurements were performed. The authors also acknowledge the financial support provided by the Polish Ministry of Science and Higher Education through SPUB subsidies for financing the maintenance of the Molecular Beam Epitaxy (MBE) Laboratory of Semiconductor Nanostructures (31/623767/SPUB/SP/2025) and the Electron Microscopy and Nanolithography Laboratory (21/599012/SPUB/SP/2024).

\end{acknowledgement}

\begin{suppinfo} \label{Sup}

%A listing of the contents of each file supplied as Supporting Information should be included. For instructions on what should be included in the Supporting Information as well as how to prepare this material for publications, refer to the journal's Instructions for Authors.

The following files are available free of charge.
\begin{itemize}
  \item Supplementary data.pdf: Additional details on sample growth and structural characterization (including RHEED and AFM images, XRD and SEM/EDX patterns), as well as supplementary ARPES images and core-level spectra, tight- binding theoretical model and energy band simulations are provided in the Supporting Information.
\end{itemize}

\end{suppinfo}

\bibliography{biblio}
\clearpage %Page break

\section{\color{red}Graphical TOC Entry}

\begin{figure}[h]
    \includegraphics[width=\columnwidth]{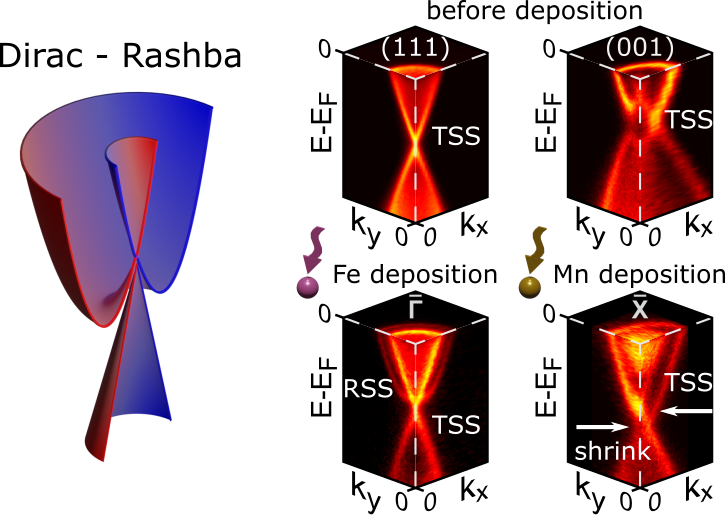}
    \label{For Table of Contents Only}
\end{figure}

\end{document}

\pagebreak

\section{Molecular beam epitaxy, RHEED caracterisation}
\label{sec:mbe}
\begin{figure*}
    \includegraphics[scale=1]{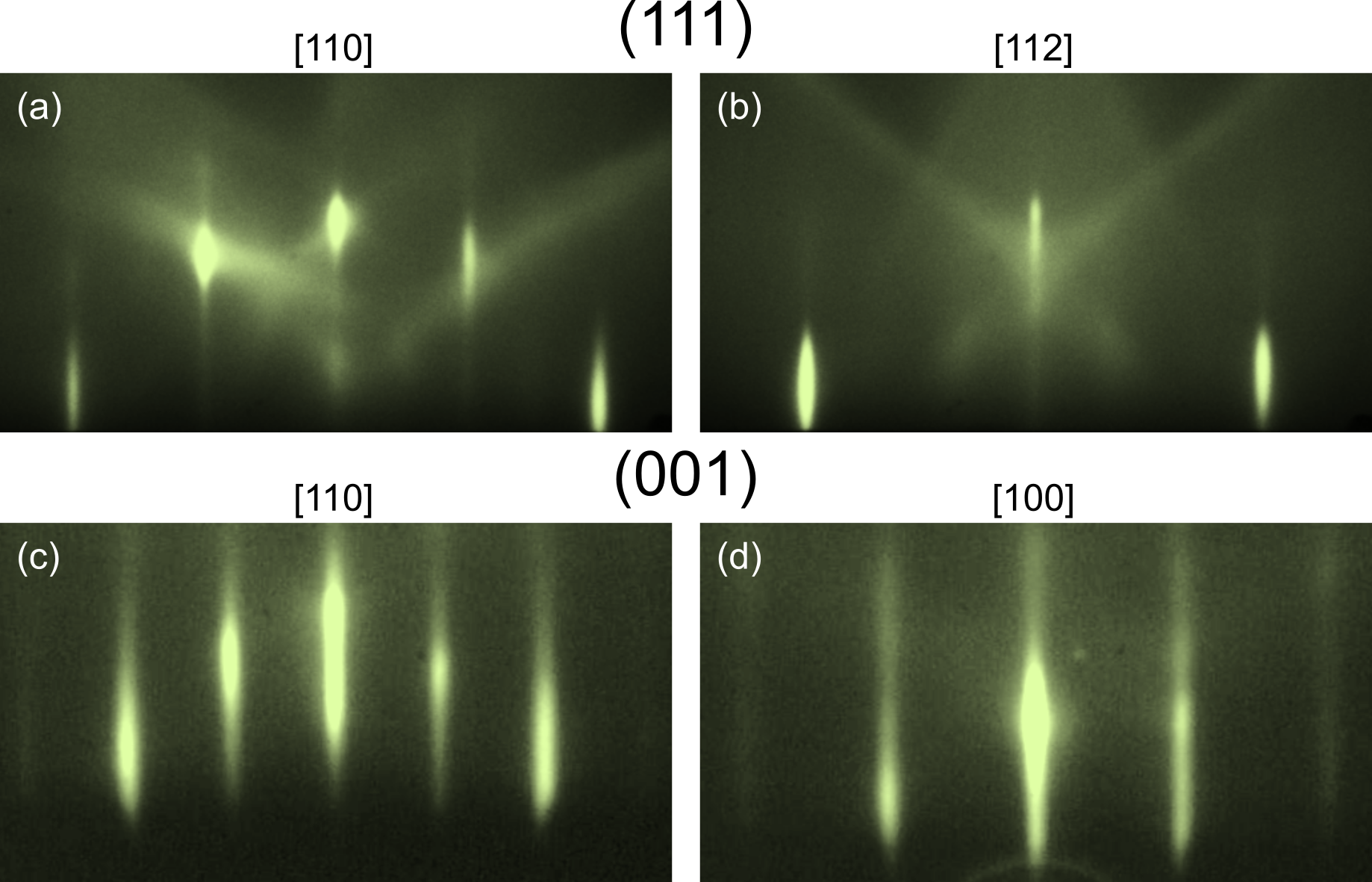}
    \caption{ Representative RHEED patterns acquired after 1~$\mu$m film growth on (a,b) (111) BaF$_2$ and (c,d) (001) KCl substrates. The corresponding azimuth directions are indicated. The presence of sharp, well-defined streaks confirms high crystalline quality and two-dimensional growth of the films.}
    \label{fig:supfigure1}
\end{figure*}

Pb$_{1-x}$Sn$_x$Se films with $x_{\mathrm{Sn}} = 0.25$ and $0.3$ were epitaxially grown by molecular beam epitaxy under ultra-high-vacuum conditions (base pressure $< 5 \times 10^{-10}$~mbar) using Riber~1000 and Veeco~GENxplor systems. 1~$\mu$m thick films were epitaxially grown on (111)~BaF$_2$ and (001)~KCl substrates using compound PbSe and SnSe effusion cells as flux sources. Bi doping at concentrations below 0.1~at.\% was achieved using a compound Bi$_2$Se$_3$ effusion cell. Film composition and doping levels were estimated from deposition rates measured with a quartz crystal microbalance positioned at the substrate location prior to experiment. The growth rate was approximately 1~\AA~s$^{-1}$. \textit{In-situ} reflection high-energy electron diffraction (RHEED) was used to monitor crystalline quality of the growing film as shown in Fig. \ref{fig:supfigure1}.

Films grown in the Riber 1000 MBE system were capped \textit{in-situ} with amorphous selenium to protect the surface from oxidation and contamination during \textit{ex-situ} transfer to the ARPES facility. Films grown in the Veeco GENxplor MBE system were transferred under ultra-high vacuum (UHV) to the SOLARIS synchrotron ARPES endstation using a Ferrovac VSN40S portable UHV suitcase (base pressure $\sim 2 \times 10^{-11}$\,mbar), maintaining vacuum integrity throughout the process.

\section{Surface characterization by atomic force microscopy}
\label{sec:surface}
Atomic force microscopy (AFM) was employed to evaluate the surface roughness and identify surface defects in as-grown films, as well as to assess possible surface degradation after transition metal deposition. This characterization is crucial for subsequent ARPES experiments.

\begin{figure*}[!h]
    \includegraphics[scale=1]{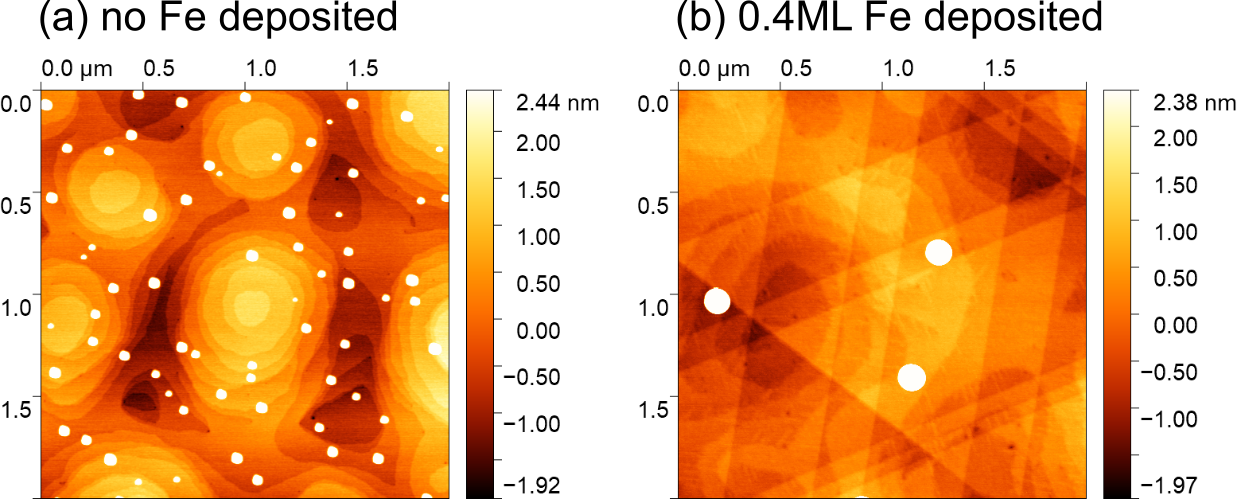}
    \caption{ Atomic force microscopy images of Pb$_{0.7}$Sn$_{0.3}$Se without (a) and with (b) Fe deposited on the surface. High-roughness features, visible as white spots on the images appear after film exposure to air.}
    \label{fig:supfigure2}
\end{figure*}

The obtained films are atomically flat, exhibiting a typical root-mean-square (RMS) roughness below 0.5 nm (Fig.~\ref{fig:supfigure2}(a,b)). At the growth temperature of 350~$^{\circ}$C used in this work, the films grow in a two-dimensional step-flow mode. This results in a characteristic spiral step structure originating from screw-type threading dislocations, which form as a result of the lattice mismatch between \PbSnSe~and BaF$_2$ (approximately 1.2--2\%)~\cite{springholz1996spiral} (see Fig.~\ref{fig:supfigure2}(a)). High-roughness features, visible as white spots in the AFM images, appear in samples that were removed from the UHV chamber and exposed to ambient conditions for several hours. These features are most likely related to the formation of selenium-rich aggregates on the surface, resulting from the outdiffusion of excess Se from within the layer. Such structures are also observed in SEM images as spherical features and can be readily re-evaporated in situ, confirming that they likely consist of volatile Se species. Deposition of submonolayer amounts of Fe does not significantly alter the surface morphology, and no metal clustering was observed on the \PbSnSe~surface (Fig.~\ref{fig:supfigure2}(b)). However, the same image (Fig.~\ref{fig:supfigure2}(b)) reveals a set of linear features intersecting at angles of 60$^{\circ}$, consistent with the threefold symmetry of the (111) surface. These features are attributed to dislocations emerging at the surface during sample cooling in ARPES experiments, caused by differences in the thermal expansion coefficients of the film and substrate at low temperatures~\cite{zogg1994thermal}. 

\section{Structure and composition. X-ray diffraction}
\label{sec:xrd}
The structural properties and composition of grown thin {\PbSnSe} films were examined by X-ray diffraction (XRD) using PANalytical X’Pert Pro MRD diffractometer equiped with a 1.6 kW x-ray tube emitting CuK$\alpha _{1}$ radiation ($\lambda$ = 1.5406~\AA), a symmetric 2 $\times$ Ge (220) monochromator and 2D Pixel detector. 

\begin{figure*}[!h]
    \includegraphics[scale=1]{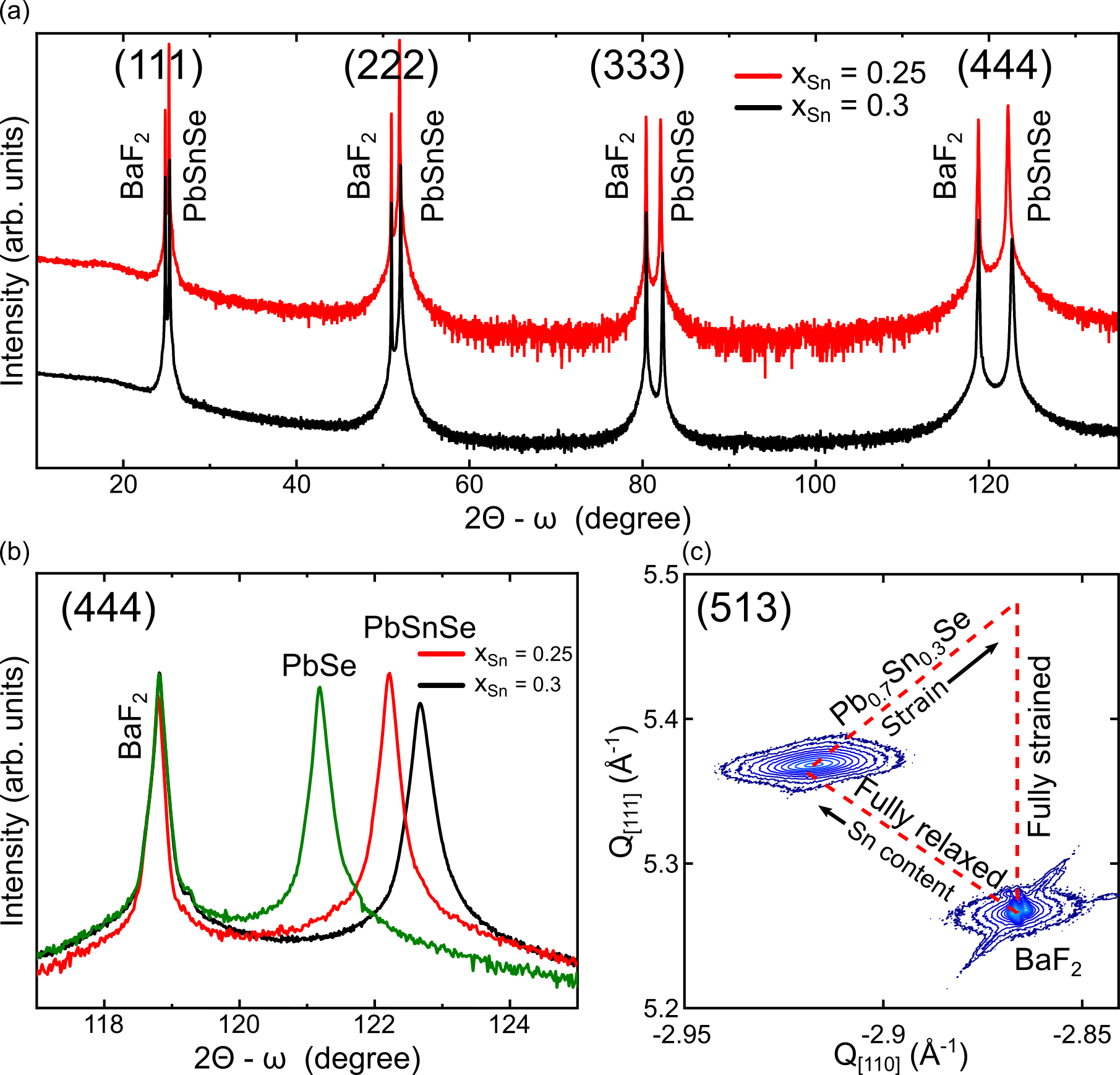}
    \caption{{X-ray diffraction of {\PbSnSe} epilayers on {\BaF} (111)}. (a) 2{$\Theta - \omega$} scan of layers with (red) x$_{Sn}$=0.25 and (black) x$_{Sn}$=0.3,  showing a single-crystaline phase of (111) oriented epilayers in both cases. (b) Zoom-in of the diffraction spectra in vicinity of (444) Bragg reflection of {\PbSnSe} and PbSe. (c) XRD reciprocal space map recorded around the asymmetric (513) reflection at room temperature, showing negligible strain induced by growth on \BaF .}
    \label{fig:supfigure3}
\end{figure*}

For both compositions ($x_{\mathrm{Sn}} = 0.25$ and $x_{\mathrm{Sn}} = 0.3$), the 1~\textmu m thick films exhibit a single-crystalline, single-phase structure with no noticeable peak broadening and asymmetry associated with alloying compared to pure PbSe, as shown in Fig.~\ref{fig:supfigure3} (a,b). An increase of the Sn content in {\PbSnSe}  leads to a shift of the diffraction peaks toward higher angles, as illustrated in Fig.~\ref{fig:supfigure3}(b), which presents the spectra in the vicinity of the (444) Bragg reflection. This shift corresponds to a decrease in the lattice constant $a{_\mathrm{PbSnSe}}$. The Sn content and the lattice constant are related by the empirical equation~\cite{Krizman2018}:

\begin{equation}
 x_{Sn} = \frac{6.1240 - a_{PbSnSe} } {0.1246},
\end{equation}
where 6.1240~\AA~is the lattice constant of pure PbSe.
The compositions of the grown films determined by XRD are in agreement within 1\% with values obtained from \textit{in-situ} beam flux measurements using a quartz crystal microbalance and from energy-dispersive analysis (see below).
Despite the relatively large lattice mismatch between the film and the substrate ($\Delta a/a$ = 1.2–2\%), the films are found to be nearly fully relaxed. XRD reciprocal space maps (RSM) measured around the asymmetric (513) reflection at room temperature (Fig.~\ref{fig:supfigure3}(c)) reveal only negligible in-plane tensile strain, below 0.1\%.

\section{Composition. Energy-dispersive X-ray spectroscopy}
\label{sec:edx}
Energy-dispersive X-ray spectroscopy (EDX) measurements were performed using a Zeiss Auriga CrossBeam Neon 40 scanning electron microscope (SEM) equipped with a QUANTAX 400 Bruker EDX system. The measurements were carried out on samples grown under conditions similar to those used for ARPES experiments. 

\begin{figure*}[!h]
    \includegraphics[scale=1]{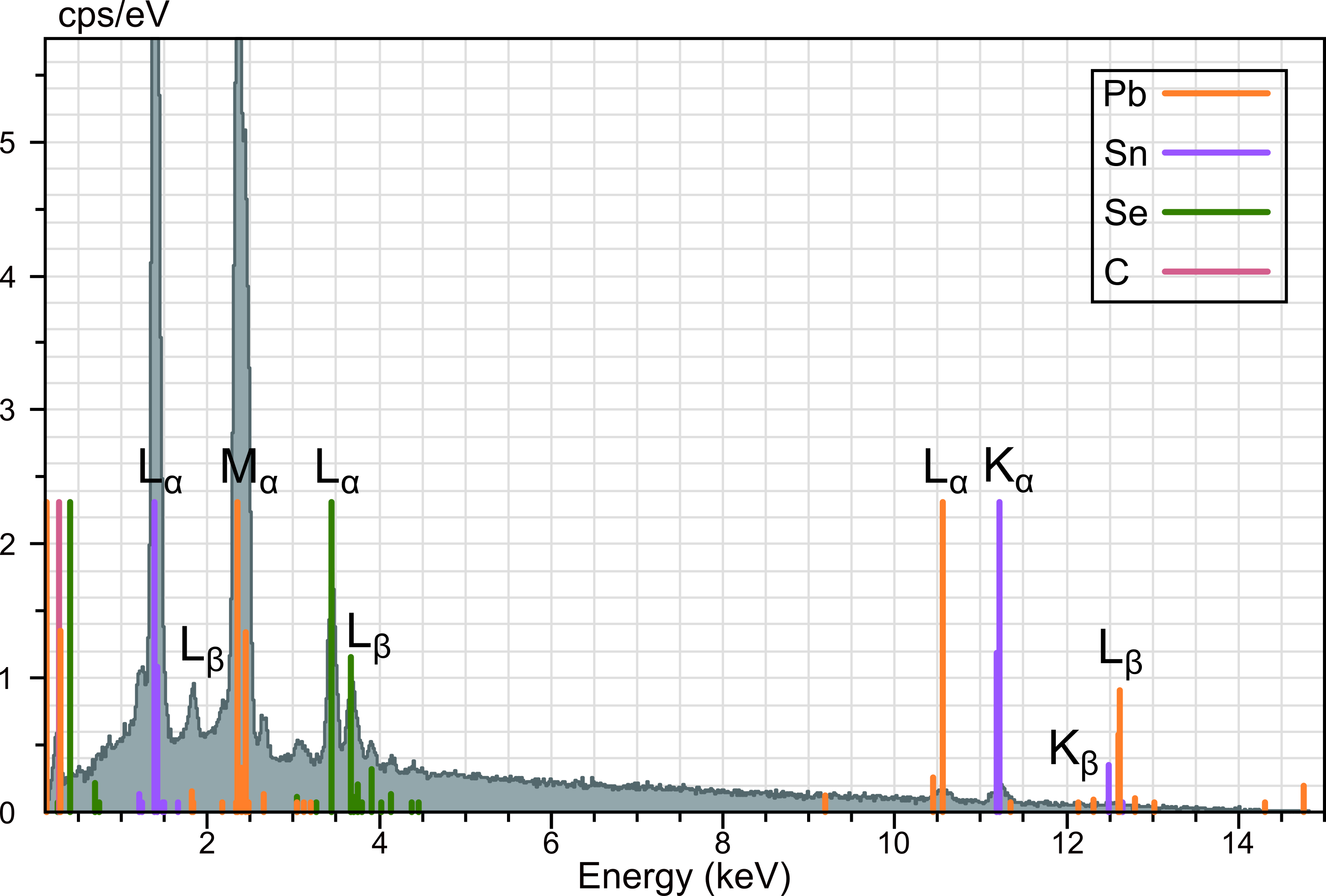}
    \caption{EDX spectrum of {\PbSnSe} with x$_{Sn}$=0.31. Presence of carbon, commonly observed in SEM measurements, are also detected.}
    \label{fig:supfigure4}
\end{figure*}

\begin{table}[!h]
\begin{tabular}{ c c }
\hline
Element & Concentration [at. \%]\\
\hline
Pb & 32.98 \\
Sn & 14.79 \\
Se & 52.23 \\
\hline
\end{tabular}
\caption{EDX composition of \PbSnSe}
\label{table:tbl1}
\end{table}

 Fig.~\ref{fig:supfigure4} shows presence of Pb, Sn and Se, along with carbon. The carbon signal is typically observed in SEM/EDX measurements for most samples introduced into the instrument, regardless of their composition. The quantitative composition obtained from EDX is summarized in Table~\ref{table:tbl1}. The Sn content is calculated as $x_{\mathrm{Sn}} = \frac{at_{\mathrm{Sn}}}{at_{\mathrm{Pb}} + at_{\mathrm{Sn}}}$ and, for the data presented in Table~\ref{table:tbl1}, yields $x_{\mathrm{Sn}} = 0.31$. This value is in agreement within 1\% with the composition determined from \textit{in-situ} beam flux measurements using a quartz crystal microbalance, as well as from X-ray diffraction.

\section{Additional angle-resolved photoemission data}

\subsection{Topological - normal insulator transition} 

The temperature-driven topological – normal insulator transition in Pb$_{1-x}$Sn$_x$Se is an important factor in assessing the suitability of these samples for further systematic ARPES studies. This transition originates from band inversion at the L points of the Brillouin zone and depends sensitively on both Sn concentration and temperature. In epilayers, the transition temperature may deviate from bulk values due to residual strain and finite-size effects.

\begin{figure*}
    \includegraphics[scale=1]{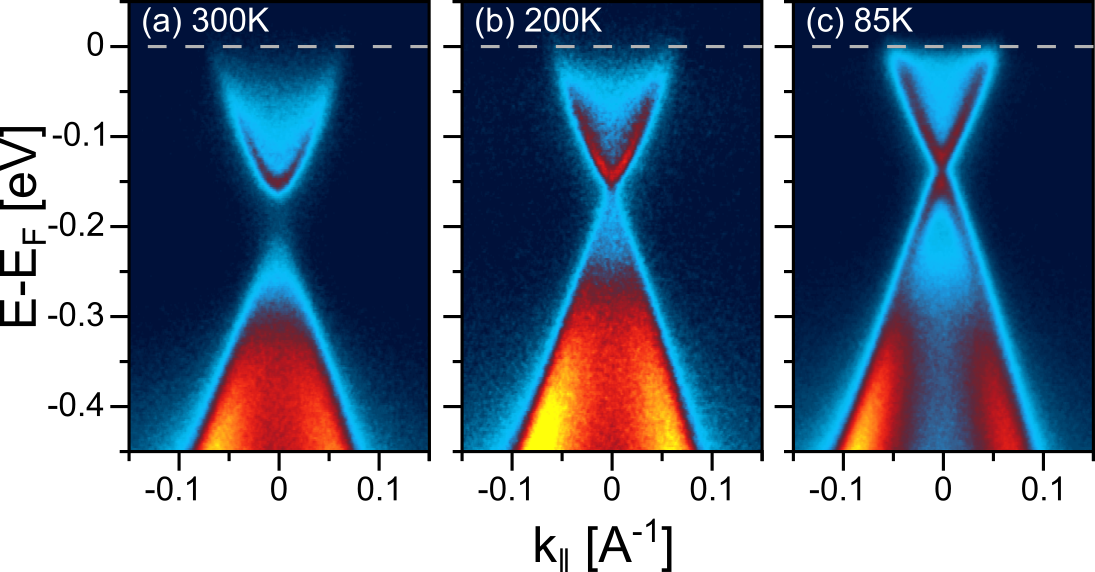}
    \caption{Temperature-driven topological phase transition in Pb$_{0.7}$Sn$_{0.3}$Se. Decreasing temperature leads to the closure of the  band gap and a transition from the trivial insulating phase (a) to the topological crystalline insulator phase (b), (c). ARPES spectra were measured at E$_{ph}$ = 18 eV in vicinity of the \Gbar~point.}
    \label{fig:supfigure5}
\end{figure*}

The evolution of the ARPES spectra at the \Gbar~point of the Pb$_{0.7}$Sn$_{0.3}$Se sample with decreasing temperature is shown in Fig.~\ref{fig:supfigure5}. At room temperature (Fig.~\ref{fig:supfigure5}(a)), the system is in the trivial (normal insulating) phase, characterized by a finite band gap of approximately 68~meV with normal band ordering. Upon cooling, the band gap decreases and eventually closes as the system approaches the topological phase transition. According to the composition- and temperature-dependent band gap relation proposed by Krizman \textit{et al.}~\cite{Krizman2018}, the transition temperature for Pb$_{1-x}$Sn$_x$Se with $x{_{Sn}} = 0.3$ is approximately 218.5 K. In our measurements, the gap is already closed at 200 K, indicating that the system is at or very close to the transition point. Further cooling to 85 K drives the system deeper into the inverted regime, where the band ordering is reversed and the material is well within the topological crystalline insulator phase.

Determining the transition temperature in the studied epilayers is essential for ensuring a consistent interpretation of the ARPES spectra and the emergence of topological surface states, in agreement with previously reported results.

\subsection{{\Mbar}-point ARPES spectra}

\begin{figure*}
    \includegraphics[scale=1]{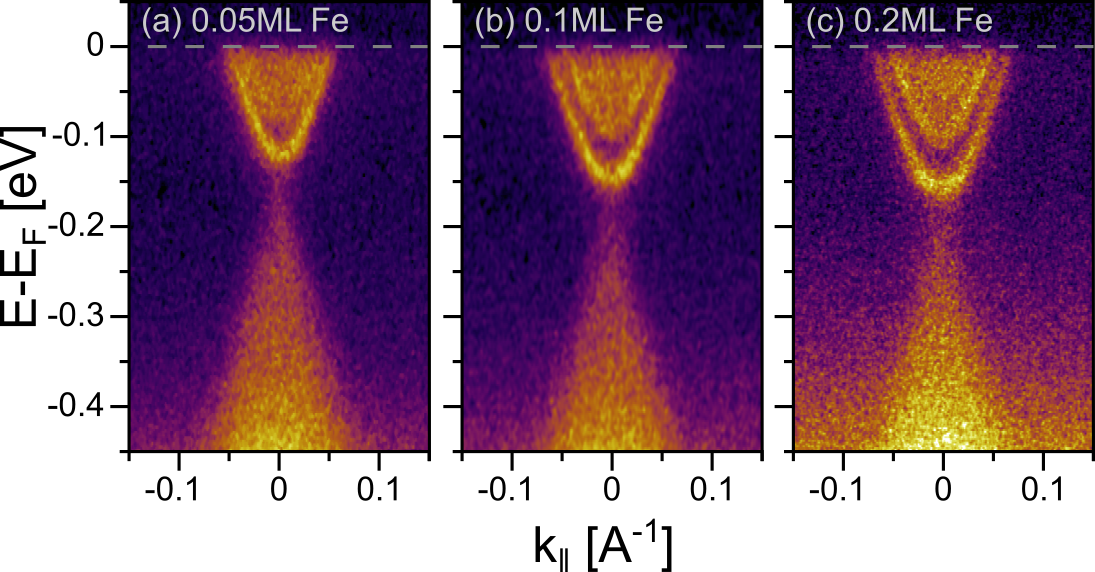}
    \caption{ARPES spectra in the vicinity of \Mbar~- point for Pb$_{0.65}$Sn$_{0.25}$Se as amount of iron deposited on the sample surface increases (a-c) measured at 77K with E$_{ph}$ 17.2eV.}
    \label{fig:supfigure6}
\end{figure*}

Typical ARPES spectra acquired near the {\Mbar} point for (111)-oriented \PbSnSe~ film obtained after Fe deposition are shown in Fig.~\ref{fig:supfigure6}(a–c)). Due to the relatively low photoemission intensity at this point of the surface Brillouin zone, the spectra are of lower quality than those obtained at the \Gbar~point and were therefore not analyzed in detail. Nevertheless, the main features of the electronic structure can still be identified. 

As expected, a Dirac-like dispersion is observed, with the topological surface state crossing within the bulk band gap. Increasing Fe coverage leads to a systematic shift of the Dirac point toward larger binding energies, consistent with the behavior observed at the \Gbar~point. The valence band shows an approximately linear dispersion - characteristic of topological surface states - while the conduction band exhibits a more parabolic dispersion with two relatively sharp states, which can indicate formation of a two-dimensional Rashba electron gas (2DEG).

 Although such states may be associated with Rashba-type splitting, no clear Kramers degeneracy point is resolved in the presented data. Consequently, no detailed analysis of the electronic structure was performed for this high-symmetry point.

\subsection{Influence of surface deposition on ARPES spectra}

\begin{figure*}
    \includegraphics[scale=1]{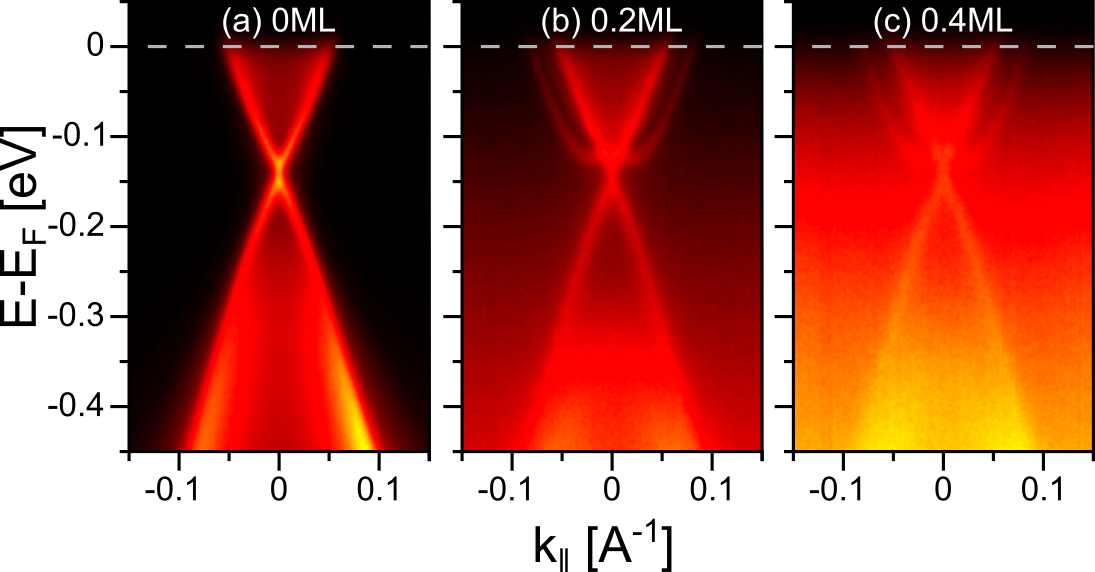}
    \caption{Influence of Fe deposition on ARPES spectra in the vicinity of \Gbar~point at 80K and for E$_{ph}$ = 18 eV. As surface roughness increases due to the deposition of transition metal, the quality and resolution of the ARPES spectra deteriorate. The spectra are shown without background subtraction.} 
    \label{fig:supfigure7}
\end{figure*}

Raw ARPES spectra (i.e., without background subtraction) obtained near \Gbar~point are shown in Fig.~\ref{fig:supfigure7} (a–c), corresponding to spectra (j–l) in Fig.~2 of the main text. As the amount of deposited Fe increases, the background noise becomes more pronounced. This behavior suggests increasing
surface roughness and the formation of a disordered transition metal overlayer. However, despite of the increasing noise, indicating a gradual degradation of surface quality, the spectral features associated with Dirac-Rashba spectrum is clearly seen.

To further assess the nature of the deposited layer, complementary RHEED measurements were performed during transition metal deposition (see ref. \cite{TUROWSKI2023}). A decrease in the intensity of the specular spot, accompanied by an increase in diffuse background, is observed, while no additional diffraction features (such as spots or rings) appear. This indicates that the deposited transition metal does not form an ordered crystalline layer, but rather a disordered, amorphous metal phase on the surface of Pb$_{1-x}$Sn$_x$Se.

\subsection{Photon energy dependence}

\begin{figure*}
    \includegraphics[scale=1]{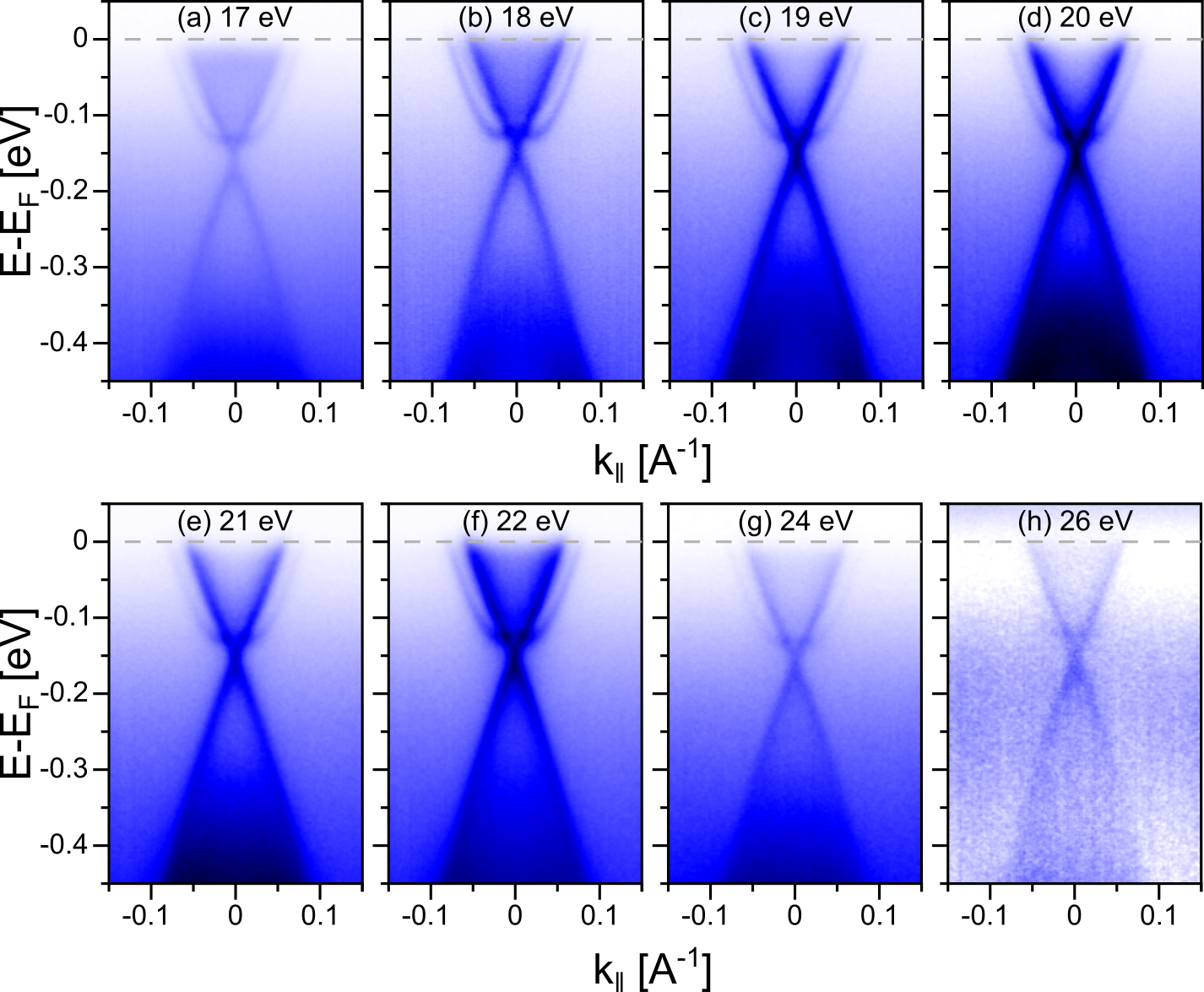}
    \caption{Photon energy dependence of Pb$_{0.7}$Sn$_{0.3}$Se with 0.2ML Fe deposited on the surface. Surface states do not disperse when the energy of the incident photon is varied.}
    \label{fig:supfigure8}
\end{figure*}

Unlike bulk electronic states, which disperse with the out-of-plane momentum component $k_{\perp}$, surface states are expected to exhibit negligible dependence on the incident photon energy $E_{\mathrm{ph}}$. To verify the surface character of the observed states, photon-energy-dependent ARPES measurements were performed on the sample with 0.2 ML of Fe deposited on the surface. The photon energy was varied between 17 and 26 eV, corresponding to a change in the probed $k_{\perp}$ range (see Fig. \ref{fig:supfigure8} (a-h)). Neither the topological surface states (TSS) nor the Rashba-split surface states (RSS) exhibit any measurable energy or momentum shift throughout the investigated photon-energy range. This absence of dispersion confirms their two-dimensional surface character. In contrast, the spectral intensity of the surface states varies significantly with photon energy, which can be attributed to photoemission matrix-element effects that modulate the ARPES intensity without altering the underlying electronic structure. In particular, the RSS are strongly suppressed in the spectrum acquired at $E_{\mathrm{ph}} = 26$ eV. Nevertheless, enhanced-intensity segments of the RSS can still be identified.

\subsection{Circular Dichroism}
\label{CD}
\begin{figure*}
    \includegraphics[scale=1]{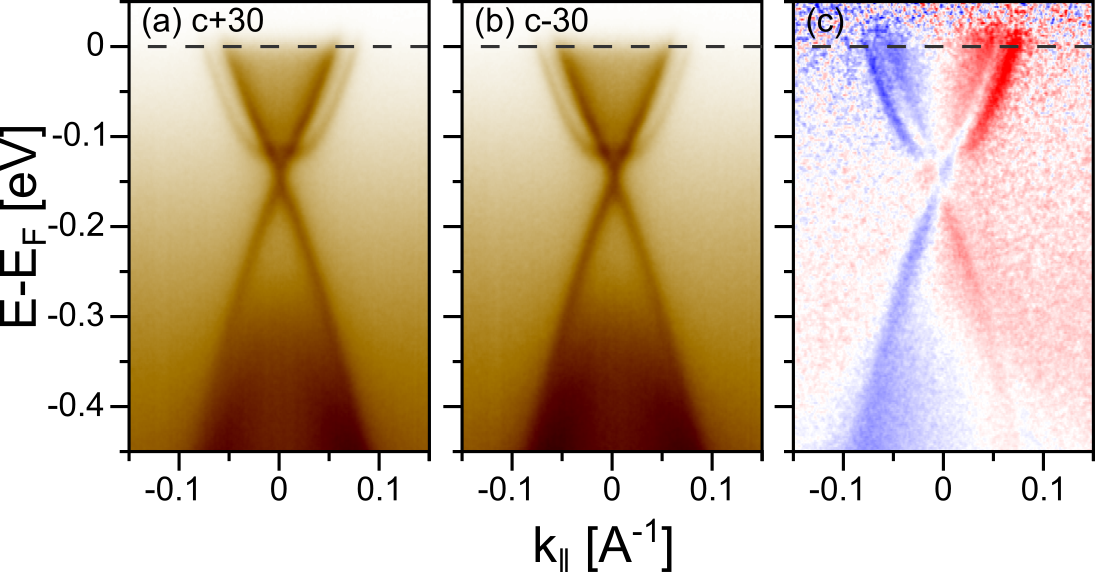}
    \caption{Circular dichroism ARPES of Pb$_{0.7}$Sn$_{0.3}$Se with 0.2ML Fe deposited on the surface. ARPES spectra measured at 80K with E$_{ph}$ 18eV in the vicinity of \Gbar~point with (a) left- and (b) right-circularly polarized light. (c) Circular dichroism, obtained as the intensity difference between panels (a) and (b), revealing a pronounced dichroic response of the surface states.}
    \label{fig:supfigure9}
\end{figure*}

Circular-dichroism ARPES measurements were performed on Pb$_{0.7}$Sn$_{0.3}$Se (111) with 0.2 ML Fe deposited on the surface. As shown in Fig.~\ref{fig:supfigure9}, clear differences are observed between spectra acquired with left- and right-circularly polarized light, resulting in a pronounced dichroic signal near the $\bar{\Gamma}$ point. The dichroism is particularly strong for the RSS, while a weaker but still discernible response is observed for the TSS. Interestingly, both the inner and outer Rashba branches exhibit the same sign of dichroism. Such behavior is not uncommon in CD-ARPES experiments, where the measured intensity asymmetry is governed by photoemission matrix elements and the orbital character of the electronic states rather than directly reflecting their spin polarization. Consequently, the sign of the dichroic signal cannot be straightforwardly related to the spin helicity of the Rashba branches.

The observed circular dichroism reflects the sensitivity of the photoemission process to the orbital texture of the surface-derived states and is consistent with the presence of strong spin--orbit coupling. While circular dichroism does not provide a direct measurement of spin polarization, the pronounced dichroic response of both the TSS and RSS supports their surface-state character. The observation of the same dichroic sign for both Rashba branches likely originates from matrix-element and final-state effects, which are known to strongly influence the circular-dichroism response in ARPES measurements.

\subsection{Core-level photoemission}

{Core level studies on \PbSnSe~samples with increasing amount of Fe deposited on the sample surface were carried out with excitation photon energy E$_{ph}$ of 90 eV at 80 K, with energy resolution of 0.05~eV. At this kinetic energy, the inelastic mean free path of the emitted photoelectrons is below 10~\AA, making the measurement highly surface sensitive and therefore well suited for investigating the initial stages of submonolayer Fe deposition.

The results are shown in Fig. \ref{fig:supfigure10}. Prior to Fe deposition, only the Pb, Sn and Se core levels are observed, confirming the high cleanliness of the sample surface. The doublet structure of the Pb~5$d$, Sn~4$d$ and Se~3$d$ spectra originates from spin--orbit splitting. The measured binding energies agree well with the reported values for PbSe and SnSe compounds \cite{shalvoy1977_CL}, reflecting the characteristic chemical environment of the constituent atoms.
The first deposition step of 0.1~ML Fe (Fig.~\ref{fig:supfigure10}, dark blue) does not produce a distinct Fe~3$p$ peak at its nominal binding energy. Instead, an additional component appears on the low-binding-energy side of the Sn~4$d$ spectrum. As the Fe coverage increases, this component gains intensity, while the intensity of the original Sn~4$d_{3/2}$ peak decreases. A similar evolution is observed for the Pb~5$d$ spectrum. At a total Fe coverage of 0.2~ML (green curves), two additional Pb~5$d$ components become clearly resolved at lower binding energies (indicated by the gray vertical lines), and their intensity further increases with increasing Fe coverage.

No isolated Fe~3$p$ peak is detected throughout the investigated coverage range. However, this does not necessarily imply the absence of Fe photoemission, since the Fe~3$p$ emission is expected to overlap with the Se~3$d$ region ((Fig.~\ref{fig:supfigure10}, gray bar)) and may additionally experience a chemical shift resulting from Fe--Se bond formation. Consistent with this interpretation, the relative intensity of the Se~3$d$ signal increases with respect to the Pb and Sn core levels after Fe deposition.
A similar behaviour has previously been reported by Scholz \textit{et al.} for Fe deposited on Bi$_2$Se$_3$ \cite{scholz2012tolerance}. In that system, the Fe~3$p$ emission overlaps with the Se~3$d$ core level, while additional low-binding-energy Bi components were attributed to the formation of Fe-induced surface bonds. By analogy, the new components observed here are most naturally explained by changes in the local chemical environment of surface Pb and Sn atoms caused by Fe adsorption. For a Se-terminated \PbSnSe\ surface, Fe is expected to bond preferentially with surface Se atoms, thereby modifying the bonding configuration of the neighbouring Pb--Se and Sn--Se bonds. Such changes in local coordination give rise to chemical shifts of the substrate core levels.
The observed shifts are consistent with this picture. The additional Pb~5$d$ components are displaced by approximately 0.74~eV with respect to the bulk-related peaks, comparable to the surface-core-level shifts reported for Bi$_2$Se$_3$ \cite{scholz2012tolerance}. The additional Sn component is shifted by approximately 1.03~eV. Since the spin--orbit splitting of the Sn~4$d_{5/2}$ and Sn~4$d_{3/2}$ levels is close to 1~eV, the shifted Sn~4$d_{3/2}$ component overlaps with the original Sn~4$d_{5/2}$ peak, explaining why the apparent intensity of the latter remains nearly constant, whereas the intensity of the original Sn~4$d_{3/2}$ peak decreases with increasing Fe coverage.
Finally, it is important to distinguish the observed chemical shifts from electrostatic shifts caused by band bending. Band bending produces an almost rigid displacement of all core levels by the same energy owing to the change in the surface electrostatic potential. In contrast, the additional components observed here are element-specific and arise only after Fe adsorption, indicating that they originate predominantly from changes in the local chemical environment associated with Fe-induced surface bonding rather than from electrostatic band bending alone.
 
\begin{figure*}
    \includegraphics[scale=1]{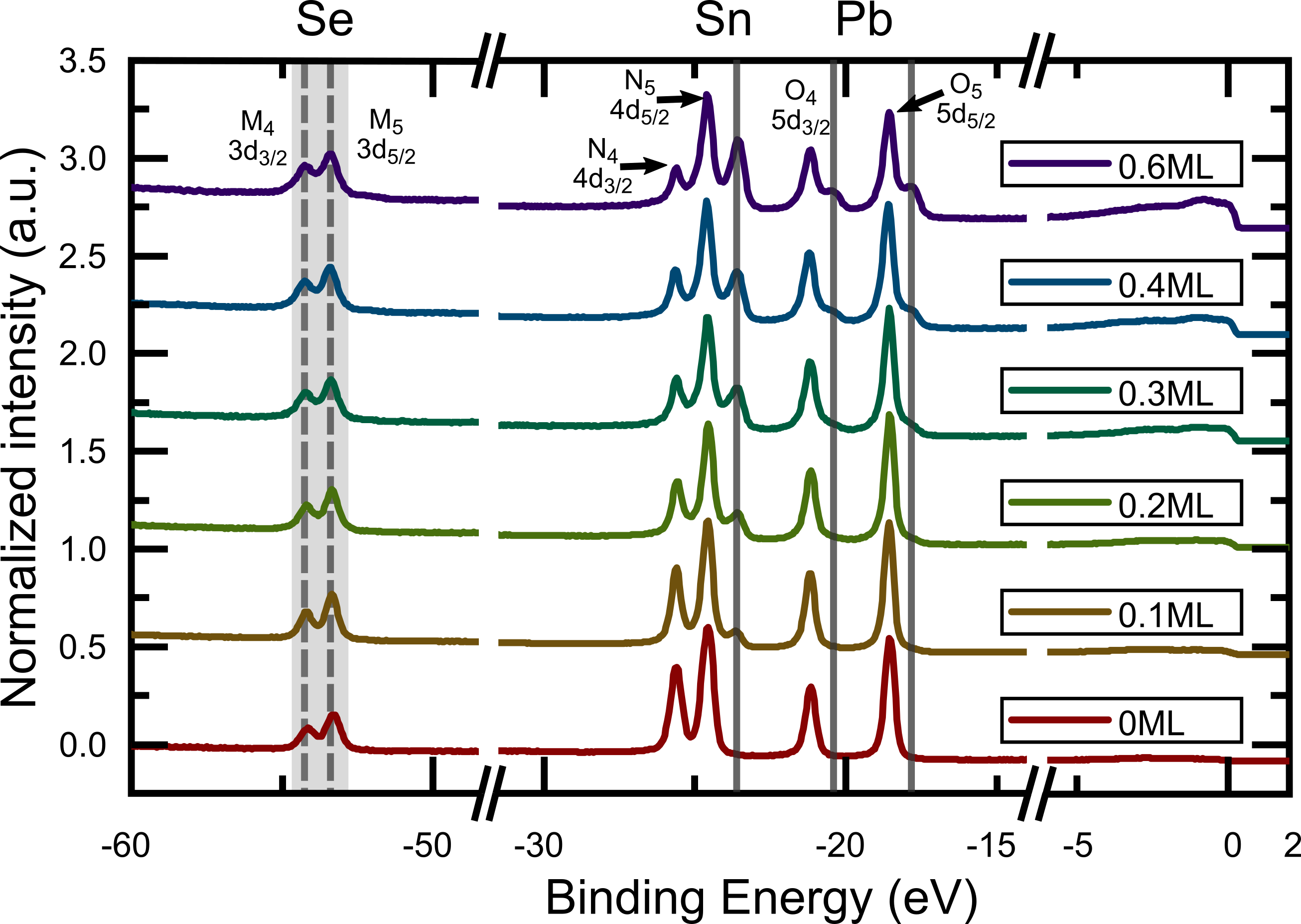}
    \caption{Photoemission spectra acquired with a photon energy of 90 eV for increasing Fe coverage on the \PbSnSe\ surface. Besides the Pb~5$d$, Sn~4$d$, and Se~3$d$ core levels, additional components (solid gray lines) emerge with increasing Fe deposition. The Fe-induced components are shifted by approximately 1.0 eV for Sn and 0.74 eV for the Pb~5$d$ doublet relative to the pristine surface peaks. Gray dashed lines and bar indicate the evolution of the Se~3$d$ doublet upon Fe deposition}
    \label{fig:supfigure10}
\end{figure*}

\section{Tight-binding model and evolution of the surface spectra}
\label{theory}

\begin{figure*}
    \includegraphics[scale=0.4]{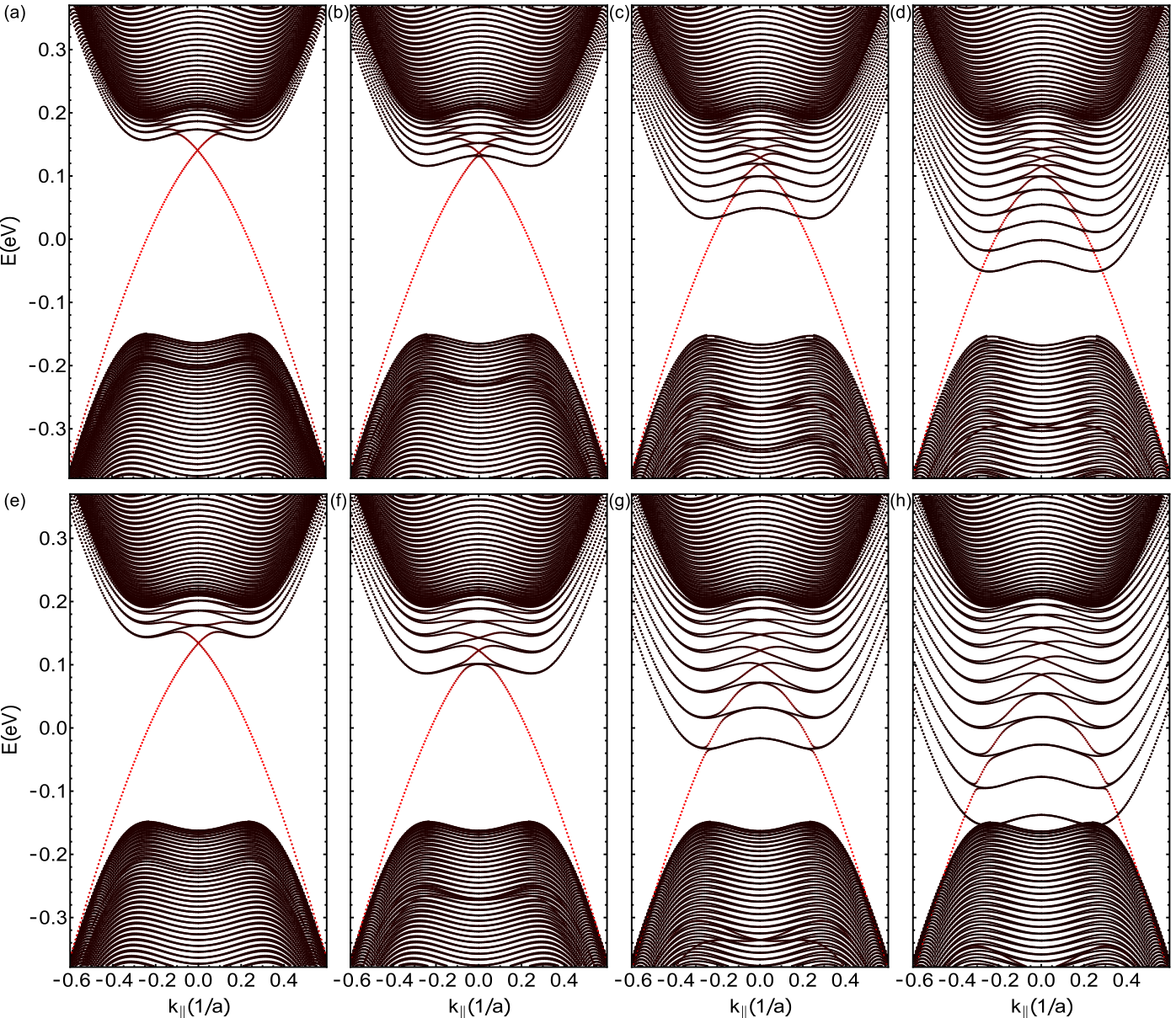}
      \caption{Energy bands of the TB model with open surfaces in the presence of the surface potential with $\xi = 10$ and: (a-d) $\lambda_{\rm TF}=50$, (e-h) $\lambda_{\rm TF}=30$  (all lengths in unit cells). Magnitude of the surface potential is increased for the two rows of plots going from left to the right as: (a) $\eta = \eta_0/2$, (b) $\eta = \eta_0$, (c) $\eta = 2\eta_0$, (c) $\eta = 3\eta_0$ with $\eta_0=0.08$ and (e) $\eta = \eta_1/2$, (f) $\eta = \eta_1$, (g) $\eta = 2\eta_1$, (h) $\eta = 3\eta_1$ with $\eta_1=0.24$. Brightness of the red color indicate magnitude of the projection of a state on a top surface of the slab, states localized at bottom surface were removed.}
    \label{fig:supfigure11}
\end{figure*}

Defining the unit cell as two atoms at positions $(0,0,0)$ and $(0,0,1)$ (taking one interatomic distance as a length unit) and the lattice translation vectors as $\vec{a}_1=(1,0,1)$, $\vec{a}_2=(0,1,1)$ and $\vec{a}_3=(0,0,2)$ we find that the three-dimensional bulk Hamiltonian can be represented in momentum space as \cite{Brzezicki2019},
\begin{eqnarray}
{\cal H}(\vec{k})&=&m\mathbbm{1}_2\!\otimes\!\mathbbm{1}_3\!\otimes\! \tau_z+t_{12}\!\!\!\sum_{\alpha=x,y,z}\!\!\mathbbm{1}_2
\!\otimes\!\left(\mathbbm{1}_3\!-\!L_{\alpha}^{2}\right)\!\otimes\! h_{\alpha}^{(1)} (\vec{k})\nonumber\\
&+&t_{11}\sum_{\alpha\not=\beta}\mathbbm{1}_2\!\otimes\!\left(\mathbbm{1}_3\!-\!\tfrac{1}{2}\left(L_{\alpha}\!+\!\varepsilon_{\alpha\beta}L_{\beta}\right)^{2}\right)
\!\otimes\! h_{\alpha\beta}^{(2)} (\vec{k})\nonumber\\
&+&\sum_{\alpha=x,y,z} \lambda \sigma_{\alpha}\!\otimes L_{\alpha}\otimes\mathbbm{1}_2,
\label{eq:suppeq2}
\end{eqnarray}

where $\vec{k}=(k_1,k_2,k_3)$, $\varepsilon_{\alpha\beta}$ is a Levi-Civita symbol, $L_{\alpha}=-i\varepsilon_{\alpha\beta\gamma}$ are the $3\times 3$ angular momentum $L=1$ matrices and spin-orbit coupling is given by $\lambda$. The mass difference between the two sites of the unit cell is encoded in a pseudospin $\tau_z$ Pauli matrix and matrices $h_{\alpha}^{(1)}(\vec{k})$ and $h_{\alpha\beta}^{(2)}(\vec{k})$ describe hopping between nearest neighbors,
\begin{eqnarray}
h_x^{(1)}&=&[\cos k_1+\cos (k_1-k_3)]\tau_x+[-\sin k_1+\sin (k_1-k_3)]\tau_y, \nonumber\\
h_y^{(1)}&=&[\cos k_2+\cos (k_2-k_3)]\tau_x+[-\sin k_2+\sin (k_2-k_3)]\tau_y, \nonumber\\
h_z^{(1)}&=&[1+\cos k_3]\tau_x-\sin k_3\tau_y,
\end{eqnarray}
and next-nearest neighbors
\begin{eqnarray}
h_{xy}^{(2)}=2\cos (k_1+k_2-k_3)\tau_z,&\quad&
h_{yx}^{(2)}=2\cos (k_1-k_2)\tau_z,\nonumber\\
h_{xz}^{(2)}=2\cos k_1\tau_z,&\quad&
h_{zx}^{(2)}=2\cos (k_1-k_3)\tau_z,\nonumber\\
h_{yz}^{(2)}=2\cos k_2\tau_z,&\quad&
h_{zy}^{(2)}=2\cos (k_2-k_3)\tau_z.
\end{eqnarray}
The multilayer system composed of $N_{\text{L}}$ $(111)$ layers can be obtained from ${\cal H}(\vec{k})$ by replacing quasimomenta $k_3$ by a real-space hopping matrix structure,
\begin{equation}
{\cal H}_{(1,1,1)}(k_{1},k_{2})=\begin{pmatrix}{\cal H}_{in} & {\cal H}_{out} & 0 & 0 & 0\\
{\cal H}_{out}^\dagger & {\cal H}_{in} & {\cal H}_{out} & 0 & 0\\
0 & {\cal H}_{out}^\dagger & {\cal H}_{in} & \ddots & 0\\
0 & 0 & \ddots & \ddots & {\cal H}_{out}\\
0 & 0 & 0 & {\cal H}_{out}^\dagger & {\cal H}_{in}
\label{Hmat}
\end{pmatrix},
\end{equation}
where diagonal blocks are given by
\begin{eqnarray}
{\cal H}_{in}(k_1,k_2)&=&m\mathbbm{1}_2\!\otimes\!\mathbbm{1}_3\!\otimes\! \tau_z+t_{12}\!\!\!\sum_{\alpha=x,y,z}\!\!\mathbbm{1}_2
\!\otimes\!\left(\mathbbm{1}_3\!-\!L_{\alpha}^{2}\right)\!\otimes\! h_{\alpha,in}^{(1)} (k_1,k_2)\\
&+&t_{11}\sum_{\alpha\not=\beta}\mathbbm{1}_2\!\otimes\!\left[\mathbbm{1}_3\!-\!\tfrac{1}{2}\left(L_{\alpha}\!+\!\varepsilon_{\alpha\beta}L_{\beta}\right)^{2}\right]
\!\otimes\! h_{\alpha\beta,in}^{(2)} (k_1,k_2) +\sum_{\alpha=x,y,z} \lambda \sigma_{\alpha}\!\otimes L_{\alpha}\otimes\mathbbm{1}_2,\nonumber
\end{eqnarray}
and off-diagonal ones by
\begin{eqnarray}
{\cal H}_{out}(k_1,k_2)&=&t_{12}\!\!\!\sum_{\alpha=x,y,z}\!\!\mathbbm{1}_2
\!\otimes\!\left(\mathbbm{1}_3\!-\!L_{\alpha}^{2}\right)\!\otimes\! h_{\alpha,out}^{(1)} (k_1,k_2)\\
&+&t_{11}\sum_{\alpha\not=\beta}\mathbbm{1}_2\!\otimes\!\left[\mathbbm{1}_3\!-\!\tfrac{1}{2}\left(L_{\alpha}\!+\!\varepsilon_{\alpha\beta}L_{\beta}\right)^{2}\right]
\!\otimes\! h_{\alpha\beta,out}^{(2)} (k_1,k_2).\nonumber
\end{eqnarray}
The matrices describing hopping are now given by
\begin{eqnarray}
h_{x,in}^{(1)}=\cos k_1\tau_x-\sin k_1\tau_y,&\quad&
h_{x,out}^{(1)}=\tfrac{1}{2}e^{-ik_1}\tau_x+\tfrac{i}{2}e^{-ik_1}\tau_y, \nonumber\\
h_{y,in}^{(1)}=\cos k_2\tau_x-\sin k_2\tau_y,&\quad&
h_{y,out}^{(1)}=\tfrac{1}{2}e^{-ik_2}\tau_x+\tfrac{i}{2}e^{-ik_2}\tau_y, \nonumber\\
h_{z,in}^{(1)}=\tau_x,&\quad&
h_{z,out}^{(1)}=\tfrac{1}{2}\tau_x+\tfrac{i}{2}\tau_y,
\end{eqnarray}
for the nearest neighbors and for the next-nearest neighbors the only non-vanishing matrices are
\begin{eqnarray}
h_{yx,in}^{(2)}=2\cos(k_1-k_2)\tau_z,&\quad&
h_{xy,out}^{(2)}=e^{-i(k_1-k_2)}\tau_z, \nonumber\\
h_{xz,in}^{(2)}=2\cos k_1\tau_z,&\quad&
h_{zx,out}^{(2)}=e^{-i k_1}\tau_z, \nonumber\\
h_{yz,in}^{(2)}=2\cos k_2\tau_z,&\quad&
h_{zy,out}^{(2)}=e^{-i k_2}\tau_z. \nonumber\\
\end{eqnarray}

\begin{figure}[t!]
    \includegraphics[scale=0.4]{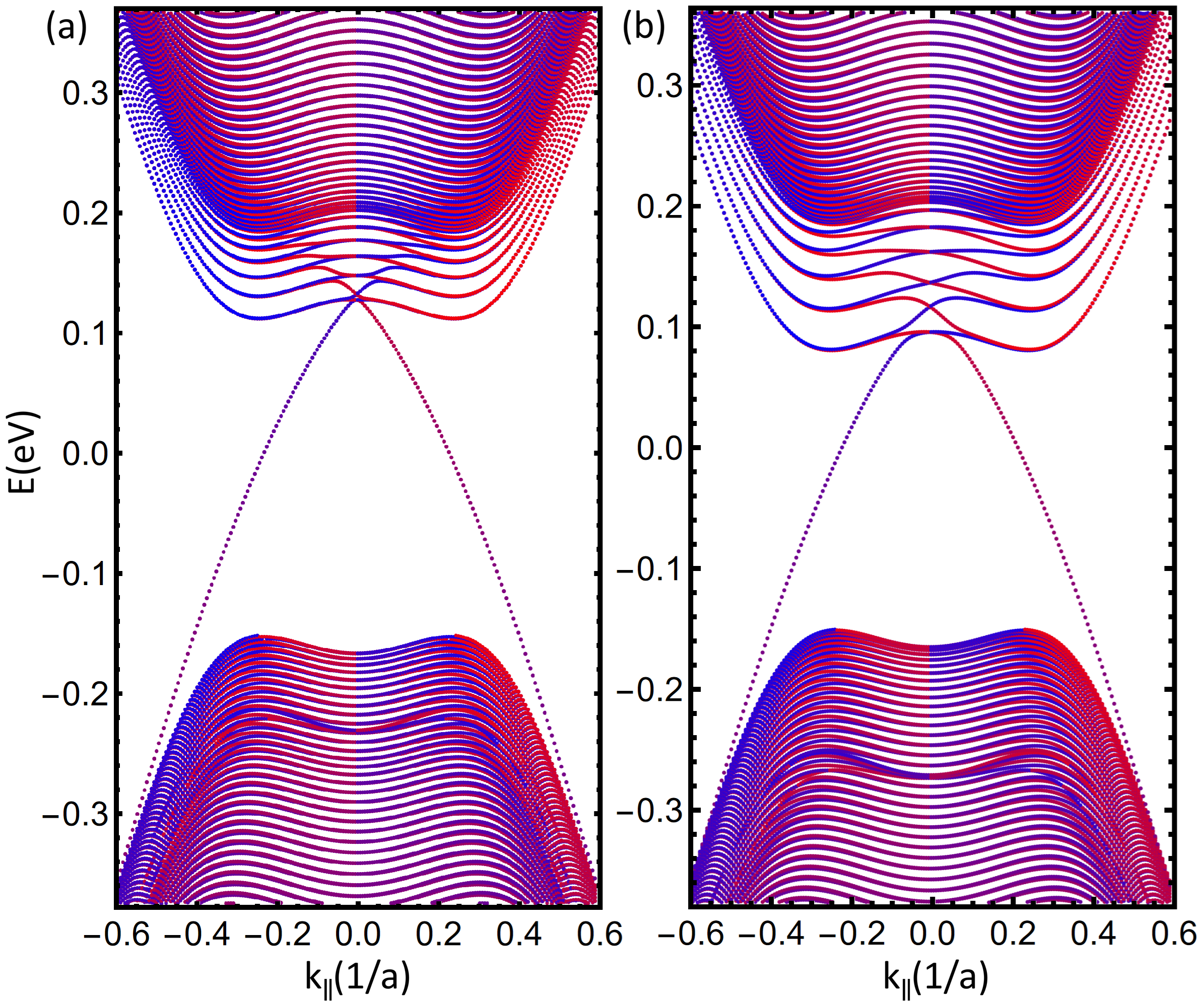}
      \caption{Energy bands of the TB model with open surfaces in presence of the surface potential with $\xi = 10$ and: (a) $\lambda_{\rm TF}=50$, $\eta = 0.08$ (b) $\lambda_{\rm TF}=30$ $\eta = 0.24$ (all lengths in unit cells). Color scale indicates the average of an in-plane component of the angular momentum $\vec{L}$, with red/blue meaning positive/negative values.}
    \label{fig:supfigure12}
\end{figure}

Additionally, we may add a surface potential term to the Hamiltonian ${\cal H}_{(1,1,1)}(k_{1},k_{2})$ in a form of
\begin{equation}
{\cal V}_{\text{surf}}={\rm diag}_N(V(N),\dots,V(2),V(1),V(0))\!\otimes\!\mathbbm{1}_2\!\otimes\!\mathbbm{1}_3\!\otimes\!\mathbbm{1}_2,
\end{equation}
where  diag$_N(d_1,\dots,d_N)$ means a diagonal matrix with entries given by $d_1,\dots,d_N$. Finally, the orthogonal surface quasimomenta are defined as
\begin{equation}
k_{x}=k_2,\quad k_{y}=\frac{1}{\sqrt 3}(2k_1-k_2).
\end{equation}
The evolution of the $(111)$ surface spectra as a function of parameters of the surface potential used in the main text is shown in Fig. \ref{fig:supfigure11}.
We see that the relative position of the Dirac cone and Rashba-split surface states can change depending on the details of the potential.

Using the approach described in the Ref. \cite{CD_L_2012} we can also relate the observed dichroic signal with the average of the atomic angular momentum operator $\vec{L}$. In Fig. \ref{fig:supfigure12} we show the energy bands for two chosen cases of the surface potential colored according to the average of the in-plane component of  $\vec{L}$. The theoretical results are similar to the observed dichroic signal, cf. Fig. \ref{fig:supfigure9}.

\pagebreak

\bibliography{biblio}